*Reliable data from low cost ozone sensors in a hierarchical network.*

Georgia Miskell[1], Kyle Alberti[2], Brandon Feenstra[4], Geoff S Henshaw[2], Vasileios Papapostolou[4], Hamesh Patel[2], Andrea Polidori[4], Jennifer A Salmond[3], Lena Weissert[1,3], David E Williams[1,*].

*Email  david.williams@auckland.ac.nz    ph +64 9 923 9877

1. School of Chemical Sciences and MacDiarmid Institute for Advanced Materials and Nanotechnology, University of Auckland, Private Bag 92019, Auckland 1142, New Zealand
2. Aeroqual Ltd, 460 Rosebank Road, Avondale, Auckland 1026, New Zealand
3. School of Environment, University of Auckland, Private Bag 92019, Auckland 1142, New Zealand
4. South Coast Air Quality Management District, 21865 Copley Drive, Diamond Bar, CA 91765, USA

**Abstract**

We demonstrate how a hierarchical network comprising a number of compliant reference stations and a much larger number of low-cost sensors can deliver reliable high temporal-resolution ozone data at neighbourhood scales. The larger than expected spatial and temporal variation of ozone in a heavily-trafficked urban environment is thereby demonstrated. The framework, demonstrated originally for a smaller scale regional network deployed in the Lower Fraser Valley, BC was tested and refined using two much more extensive networks of gas-sensitive semiconductor-based (GSS) sensors deployed at neighbourhood scales in Los Angeles: one of ~20 and one of ~45 GSS ozone sensors. Of these, ten sensors were co-located with different regulatory measurement stations, allowing a rigorous test of the accuracy of the algorithms used for off-site calibration and adjustment of low cost sensors. The method is based on adjusting the gain and offset of the low-cost sensor to match the first two moments of the probability distribution of the sensor result to that of a proxy: a calibrated independent measurement (usually derived from regulatory monitors) whose probability distribution evaluated over a time that emphasizes diurnal variations is similar to that at the test location. The regulatory measurement station physically closest to the low-cost sensor was a good proxy for most sites. The algorithms developed were successful in detecting and correcting sensor drift, and in identifying locations where geographical features resulted in significantly different patterns of ozone variation due to the relative dominance of different dispersion, emission and chemical processes. The entire network results show very large variations in ozone concentration that take place on short time- and distance scales across the Los-Angeles region. Such patterns were not captured by the more sparsely distributed stations of the existing regulatory network and demonstrate the need for reliable data from dense networks of monitors.

**Keywords:** air quality, air pollution, ozone, calibration, low-cost sensor network, maintenance

# 1. Introduction

Measurement of local-scale variations in air quality with high temporal resolution is now a topic of significant interest, which is being addressed through the development of networks of low-cost instruments. For example, although $O_3$ concentration may have a relatively regular spatiotemporal spread over large distances(Chameides et al., 1992), determined by the advection of large scale air masses such as marine air (Ainslie and Steyn, 2007; Weissert et al., 2017), small-scale variations across a city can be very large (Sadighi et al., 2018), and more correlated with nitric oxide emissions and nitrogen dioxide generation. Given the strong gradients in pollutant concentrations observed in both time and space, it is important to quantify these patterns and to elucidate the dominant processes driving them if population exposure in urban areas is to be accurately determined (Pattinson et al., 2017; Salmond et al., 2018)

As a consequence of developments in instrumentation and communications, the deployment of networks of low-cost sensors at high spatial density is now feasible. The term 'sensor' is often taken to mean just the detection element, but here, for convenience to distinguish different types of measurement instrument, the term 'sensor' refers to the assembly of the detection element, measurement electronics, air-inlet, air-sampling and communications systems, and housing and mounting that together deliver the measurement result; and 'low-cost' refers to such sensors whose installed capital cost is less than about 2% of that of a regulatory-standard reference instrument. For networks of such low-cost devices, the critical problem is the need to verify the reliability of the results with minimum physical intervention or site visits, which will dominate the network costs. Verifying reliability here means establishing a calibration that, within acceptable bounds to be defined, relates the instrument result to the otherwise unknown local concentration. Minimising cost means avoiding expensive routine on-site calibration. In principle, a Bayesian framework could be applied, using conditional probability distributions of various forms of evidence to check calibration stability of individual instruments and if necessary adjust them. However, such methods generally required large amounts of training data and are not necessarily transparent. Thus, in previous work, we described a transparent management framework that would allow use of general knowledge of the sensor and pollutant in order to detect device drift(Alavi-Shoshtari et al., 2018; Alavi-Shoshtari et al., 2013; Miskell et al., 2016; Miskell et al., 2018). Such knowledge could include diurnal patterns and geographical information such as land use as well as cross-correlations across a network(Alavi-Shoshtari et al., 2018; Alavi-Shoshtari et al., 2013; Miskell et al., 2016). We then extended these ideas to a solution to the calibration problem for low-cost air-quality sensors in networks (Miskell et al., 2018). The ideas were developed from a specification of the purpose of a low-cost network as supplementing a compliant ambient air monitoring network, extending coverage and providing reliable information for communities, including improved local coverage for exposure assessment and enhancing source compliance monitoring. Thus, the complete network is hierarchical: at the top are well-maintained, compliant instruments(Miskell et al., 2018). The definition of reliability in this context was derived from the stated purpose(Miskell et al., 2016). The ideas were developed using data from a network of low-cost $O_3$ sensors deployed around the Lower Fraser Valley (LFV) in Canada including the central urban area of Vancouver. The framework exploited network cross-correlations averaged over time, did not need large training sets to operate, was developed to work autonomously so that analytics could occur in 'real-time', and was based on transparent and simple assumptions. The concept of a proxy was introduced: that is, a reliable source of data within the network, at a different location to the site under observation, whose data have an understood expectation of probability distribution in relation to the site under observation. Miskell *et al* used land-use similarity as the criterion for determining similarity of probability distribution of pollutant concentration(Miskell et al., 2016; Miskell et al., 2018).

In the hierarchical network design, the reference stations have three roles: to provide regulatory-quality data at selected sites; to establish appropriate criteria for choice of proxies, by comparison between the reference sites; and to provide proxy data that verify the reliability of the low-cost network that is intended to extend the spatial scale (Table 1). Whilst a proxy could also, for example, be a spatio-temporal computational model (for example the comparison in Bart *et al* (Bart et al., 2014) ) this would be computationally intensive. The framework as a whole, which is set out below, involves three models, designed for transparency, clarity of purpose of each, clear comparison with instrument standards, and flexibility: a proxy model; a measurement model, within which industrial standards for low-cost instrument performance can be incorporated; and a 'semi-blind' calibration model, which also includes a decision framework.

Adapting ideas of tests for industrial process stability to determine instrument performance relative to the proxy, parameters are defined that can be tracked using control charts to detect sensor drift and distinguish this from periodic atmospheric fluctuations. The important concepts are:

1. *A proxy model*. If $X_{j,t}$ denotes the true concentration at site *j* and time *t*, $Y_{j,t}$ denotes the sensor result, and $Z_{k,t}$ the proxy site, *k*, then over some time $t_d$ that is sufficiently long to average short-term fluctuations, $Y_j$ and $Z_k$ are two different estimates of the empirical cumulative probability distribution of $X_j$ whose similarity can be tested using the Kolmogorov-Smirnov (KS) test. Using a control chart to track the time variation of the marginal probability of the KS test, $p_{KS}$, between $Y_i$ and $Z_k$ signals an alarm that $Y_i \neq X_i$ (Miskell et al., 2016).

The proxy model is defined in terms of the unknown $X_j$ evaluated over the interval $(t-t_d:t)$:

$$E\langle X_{j,t-t_d:t}\rangle = b_0 + b_1 E\langle Z_{k,t-t_d:t}\rangle + e_{j,t-t_d:t} \qquad (1)$$

$$\text{var}\langle X_{j,t-t_d:t}\rangle = b_1^2 \text{var}\langle Z_{k,t-t_d:t}\rangle + \text{var}\langle e_{j,t-t_d:t}\rangle \qquad (2)$$

where E<> denotes the mean and var<> the variance over the interval. For a 'good' proxy, $b_0$, $b_1$ and $e$ would fluctuate within defined bounds and $b_0 \approx 0$, $b_1 \approx 1$, var($e$) << var($Z$). Since $X$ is unknown, some means to check the stability of the proxy given only the measurement results of the network is required. The simplest way is to compare results across the well-calibrated reference instruments using various choices of proxy for these. The regulatory network data are used to establish appropriate proxies for the low-cost network, which in turn is used to extend the scope of the regulatory network to neighbourhood scale.

2. *A measurement model.* Sensor calibration during manufacture is assumed to establish $Y_t$ as a linear predictor of $X_t$.

$$X_{j,t} = a_0 + a_1 Y_{j,t} + \varepsilon_{j,t} \qquad (3)$$

where immediately following calibration, $a_0 \approx 0$, $a_1 \approx 1$ and the error, $\varepsilon_{j,t}$, is a zero-mean random variable, within a defined specification. Industrial standards can be set up defining acceptable bounds on the parameters of equation (3) for sensors at the point of delivery from the manufacturer. In the field, control charts of the parameters

$$\hat{a}_1 = \sqrt{\text{var}\langle Z_{k,t-t_d:t}\rangle / \text{var}\langle Y_{j,t-t_d:t}\rangle} \qquad (4)$$

$$\hat{a}_0 = E\langle Z_{k,t-t_d:t}\rangle - \hat{a}_1 E\langle Y_{j,t-t_d:t}\rangle \qquad (5)$$

test for drift of the instrument or for failure of eq (3), for example due to cross-sensitivity of the sensor signal to other, perhaps correlated pollutants(Miskell et al., 2016) . Sensor drift can be detected through drift of the estimates $\hat{a}_0$ and $\hat{a}_1$. We call this the Mean-Variance (MV) moment-matching test for intercept and slope.

3. *A "semi-blind calibration" model.* This is very simple(Miskell et al., 2018). It states that the best estimate of the unknown, $X_{j,t}$, is given by

$$\hat{X}_{j,t} = \hat{a}_0 + \hat{a}_1 Y_{j,t} \tag{6}$$

where $\hat{a}_0$ and $\hat{a}_1$ are given by equation (4) and (5). If the distributions are characterized by only two parameters, then the site distribution, $\hat{X}_{j,t-t_d:t}$ is constrained to be the same as that of the proxy, $Z$. As discussed by Miskell et al(Miskell et al., 2018), that raises the question of whether the local site information is lost. However, practical distributions are not simple 2-parameter ones though they may be similar if sufficiently averaged, and as shown by Miskell et al, using both simulated and field data, this simple procedure indeed captured the local variations, particularly the extreme values, and corrected drifting devices(Miskell et al., 2018).

Table 1 summarises the relationship between the reference network and the low-cost sensor network.

| Regulatory network | Low-cost sensor network |
|---|---|
| Well-maintained and validated; regular site calibration to regulatory standards | Factory calibrated; $X_{j,t} = a_0 + a_1 Y_{j,t}$ |
| Federal Reference Method "ground truth" data at selected sites | "indicative" method defined by industry standards. Here: $a_0$ = 0 ± 5 ppb; $a_1$ = 1 ± 0.3 (U.S. Environmental Protection Agency, 2013) |
| Determine appropriate proxies | Extends network to neighbourhood scale, to determine small-scale spatial and temporal variation |
|  | Checked and adjusted against proxy distribution, evaluated over $t_d$ and $t_f$. |

Table 1. Summary of characteristics and relationship of regulatory and low-cost sensor network

In the previous work(Miskell et al., 2016; Miskell et al., 2018), proxies were well-maintained reference stations chosen based on land-use similarity. Basing the ideas on principles of land-use regression gave a simple and effective solution. The previous work demonstrated success in both identifying and correcting sensor drift. However the transferability of the success of this model to regions of different geography, meteorology, traffic and population is unknown. Thus the question is whether the previous success was due to particular geographical features of the LFV: for example, that the valley is relatively confined, free from major geographical features within the valley itself, and has a relatively smooth $O_3$ field, so that cross-correlation between sites was high. The LFV also has an extensive network of well-maintained reference sites and good proxies were easy to identify using very general land-use similarity. Thus one key issue, which we address in the present work, is the selection of reliable proxies for a region with mixed land-use and variable geographic features.

We apply the ideas in a setting which is much more geographically variable, and demonstrate the generality through a study of two local-scale networks in different locations in Southern California, including the city of Los Angeles. The greater Los Angeles region is different from the LFV where the management framework was first devised. First, the population is much higher, with around 4.2 million inhabitants in the Inland Empire region (i.e. San Bernadino and Riverside Counties) and 9.8 million in the Los Angeles city region (U.S. Census Bureau, 2010). In comparison, the Metro Vancouver area has around 2.5 million people (Statistics Canada, 2016). Second, motor vehicle traffic in Southern California is more intense, with annual average daily traffic (AADT) estimates over 370,000 in some locations (U.S. Department of Transportation, 2016). This is in comparison to the LFV where the AADT of a major route through a large tunnel is less than 85,000 (British Columbia, 2016). Furthermore, different latitudes and seasons resulted in longer sunlight hours during the LFV network deployment (~16 h vs. ~11 h during the measurement campaign in Southern California that is reported here): sunlight is a known important precursor to $O_3$ formation(Chameides et al., 1992). Some similarities are shared between the LFV and Southern California, with sea breezes and mountain ranges affecting the regional weather patterns and pathways, causing elevated $O_3$ downwind of the urban center and away from the coast (Ainslie and Steyn, 2007; Lu and Turco, 1995; Sadighi et al., 2018). The terrain in Southern California is complex, leading to complex patterns of atmospheric convection that strongly influence surface ozone concentrations(Bao et al., 2008; Neuman et al., 2012a; Ryerson et al., 2013).

## 2. Methods

### 2.1 Study Area

The two local networks were around the Los Angeles city ("LA", *n* = 20) and the Inland Empire ("IE", *n* = 45) regions. Both regions often fail to meet attainment of the standards for $O_3$ air pollution levels set out by the US EPA (U.S. Environmental Protection Agency, 2006). In each local network, five sensors were co-located with regulatory analyzer stations, which were used for validation of the framework (Figure 1). Some of these sensors were moved for a short time to co-locate with other sensors, to illustrate the idea of a mobile device deployed as a 'buddy' to check local calibration. The IE area was recently examined by Sadighi et al. (Sadighi et al., 2018) using low-cost sensors to assess the local-scale $O_3$ variation, with evidence of significantly higher spatial variability than was captured by the regulatory network.

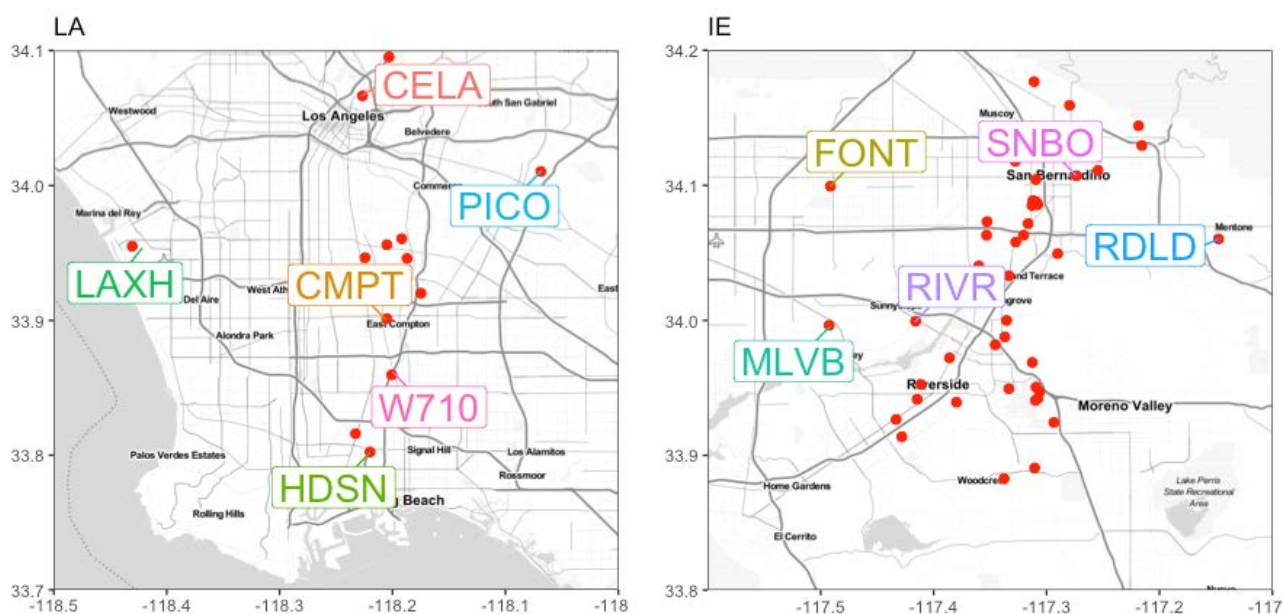

*Figure 1: Locations of the low-cost sensors in the two local-scale networks. Red points are the non-co-located sensors and those labeled sites are the sensors that are co-located with a regulatory analyzer monitoring station.*

The characteristics of the ten regulatory analyser sites are in Table 2. Each site had a Federal Equivalent Method Ozone analyzer (model 49i, Thermo Fisher Scientific, Waltham, MA and model 400E, Teledyne API, San Diego, CA) maintained and regularly serviced by the South Coast Air Quality Management District (SCAQMD). Regulatory measurement locations are selected with regard to a number of criteria, which include the highest concentration, population exposure, source impact and background. The sites had different urban surroundings and spanned from residential to industrial land-uses. However, classifying locations into land-uses was more difficult than in the LFV network due to the complexity of the urban area (e.g. highly mixed land-uses). Road traffic within both areas is high and is likely the dominant sources of precursors to $O_3$ formation ($NO_2$ and volatile organic compounds) and of nitric oxide that reacts with ozone to form nitrogen dioxide.

Table 2: Descriptions for the ten regulatory analyzer locations. Max concentration and number of days the standard was exceeded from the 2016 summary (South Coast Air Quality Management District, 2016), annual average daily traffic (AADT) within 5 km from the California Department of Transportation for 2016 traffic volumes (www.dot.ca.gov/hq/tsip/gis/datalibrary/), site types from the site survey reports(Bermudez and Fine, 2010)- and land-uses from site visits and aerial image examination. Site types: HC = highest concentration, PE = population exposure, B = background.

| AQS Name | AQS ID | Instrument ID | Latitude | Longitude | Elevation (masl) | Max. Hourly Value (ppb) | Days O₃ Standard Exceeded (> 70 ppb 8 h) | Vehicle AADT < 5 Km (000s) | Site Type | Main land-Use < 1 Km |
|---|---|---|---|---|---|---|---|---|---|---|
| Rubidoux | RIVR | 100 | 33.9995 | -117.4160 | 248 | 142 | 69 | 169 | HC | Residential |
| Mira Loma | MLVB | 101 | 33.9964 | -117.4926 | 220 | 140 | 65 | 167 | PE | Residential |
| San Bernardino | SNBO | 102 | 34.1072 | -117.2733 | 316 | 158 | 106 | 87 | HC | Residential/ Industrial |
| Fontana | FONT | 103 | 34.0994 | -117.4914 | 363 | 139 | 49 | 183 | PE | Industrial/ Commercial |
| Redlands | RDLD | 104 | 34.0604 | -117.1476 | 475 | 145 | 97 | 92 | PE | Residential |
| Pico Rivera | PICO | 161 | 34.0104 | -118.0687 | 58 | 111 | 6 | 163 | HC | Residential |
| Compton | CMPT | 166 | 33.9014 | -118.2051 | 22 | 98 | 1 | 220 | PE | Residential |
| LAX Hastings | LAXH | 176 | 33.955 | -118.4305 | 37 | 87 | 2 | 67 | PE, B | Residential/ Industrial |
| Long Beach (Hudson) | HDSN | 177 | 33.8024 | -118.2199 | 10 | 79 | 0 | 118 | PE | Industrial/ Commercial |
| Central LA | CELA | 182 | 34.0664 | -118.2266 | 89 | 103 | 4 | 169 | PE | Urban |

Data spanned from January – July 2018 for the IE network and from March – July 2018 for the LA network. Missing periods in the data were either from reference instrument calibration or from sensors going offline or from the mobile measurement campaign. The time-series for the co-located sensors showed that whilst most tracked the regulatory analyzer $O_3$ data well other sites showed a clear drift over time (Figure 2). The site co-located at Redlands (RDLD) was of particular interest: here the low-cost sensor tracked the regulatory station well, but the data from both types of sensors showed significantly higher minimum $O_3$ concentration than the other sites in the network, illustrating some of the geographical variability over the region.

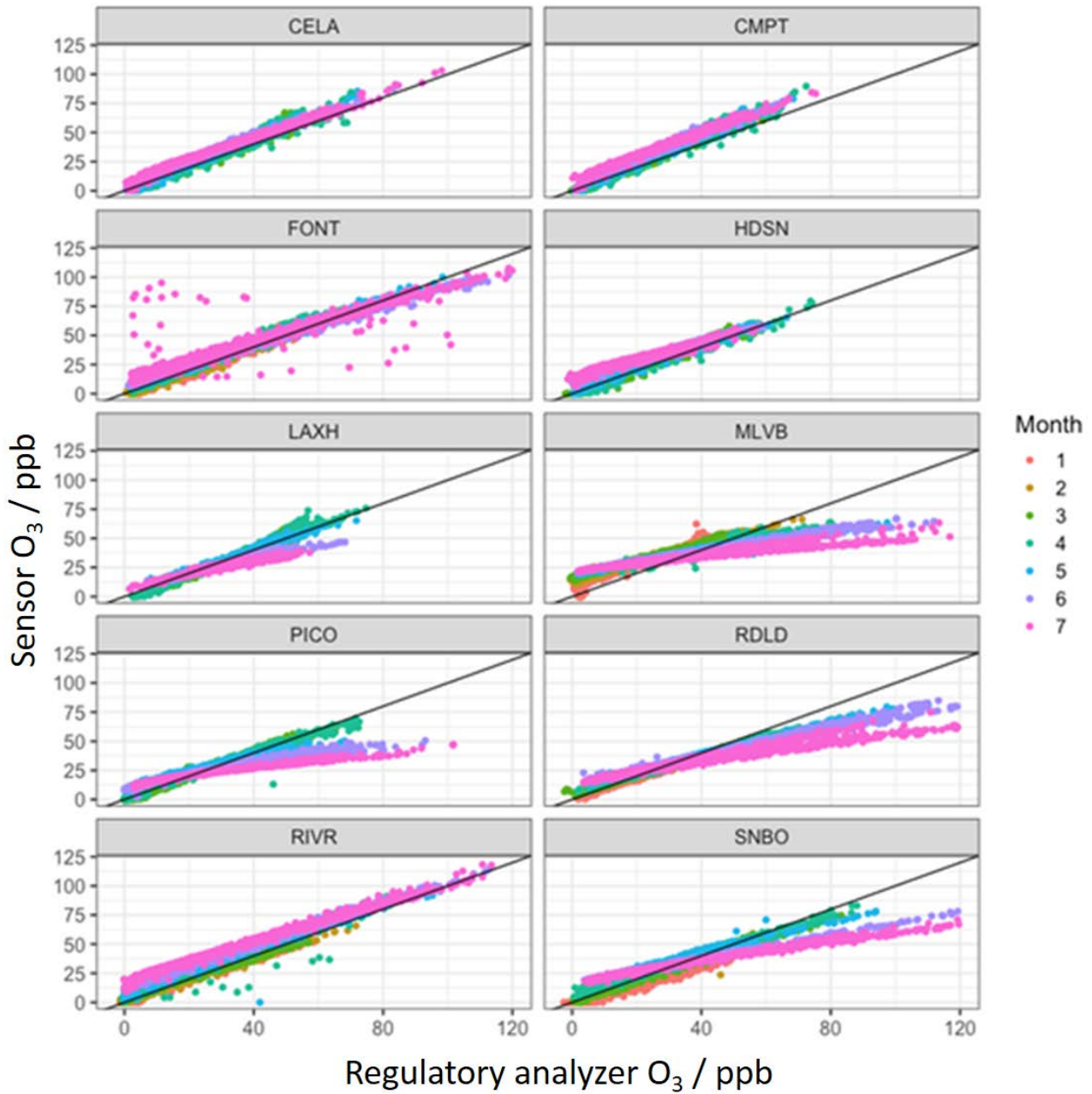

*Figure 2: Scatterplots of the raw hourly-averaged low-cost sensor data against the co-located regulatory analyzer data over the seven-month period. The line is the 1:1 line. Data are plotted separately for each month, distinguished by colour (in legend). Month 1 is January 2018 – 7 is July 2018*

## 2.2 Low-cost sensors

The deployed 'low-cost' devices are the AQY sensors from Aeroqual Ltd, Auckland, New Zealand. The $O_3$ sensor uses a gas-sensitive semiconducting (GSS) oxide, $WO_3$, as the detection element(Aliwell et al., 2001; Hansford et al., 2005; Utembe et al., 2006; Williams et al., 2002). The sensor has been extensively validated in both laboratory and field studies which have shown equation 1 to hold (with changed parameter values) even when the sensor output has drifted (Air Quality Sensor

Performance Evaluation Center, 2018; Bart et al., 2014; Cavellin et al., 2016; Lin et al., 2017; Miskell et al., 2018; Williams et al., 2013).  Here, also, the uncorrected sensor values correlated linearly throughout to the regulatory station with which each device was co-located, reflecting the fact that, despite drifts, the sensor output remained linear, as required by equation 1. The $WO_3$ detection element is insensitive to NO at typical atmospheric concentrations.  The sensor uses a combination of temperature modulation and air-flow modulation essentially to eliminate interferences due to variation of humidity or the presence of other pollutants such as $NO_2$ and volatile organic compounds at typical atmosphere concentrations(Bart et al., 2014).  Data are measured each one minute and are communicated to a server using the 4G cellular network. Key features of the sensor include solar shields to regulate heat, sophisticated inlet configuration (inert dust filters; anti-static and inert materials) and algorithms that trap known failure modes (Williams et al., 2013).

### 2.3 Management Framework

The management framework described by Miskell et al. (Miskell et al., 2016) has two important timescales: the running time over which the probability distributions are determined, $t_d$, and the timescale to determine whether a drift is a temporary excursion due to atmospheric variability or indicative of an instrument or proxy failure, $t_f$. As in Miskell et al. (Miskell et al., 2016), we chose $t_d$ = three days and $t_f$ = five days. These timescales were selected to be long enough to recover the average diurnal variations, yet short enough to allow for reasonable response times. There are three alarms: significance test using KS, $p_{KS} < 0.05$; $0.7 < \hat{a}_1 < 1.3$; $-5$ppb $< \hat{a}_0 < 5$ ppb, and an alarm is signaled when any of these conditions is maintained for a duration $> t_f$. The alarm limits for slope and offset are arbitrary and based on US EPA guidelines for indicative air quality monitoring (U.S. Environmental Protection Agency, 2013). The choice of threshold, $p_{KS}$, is subtle. If $t_d$ is made longer then there are more data points hence the statistical test of difference between distributions becomes more sensitive; however, the difference between distributions in a practical sense does not necessarily become more significant. Rather, small differences between two sets of data can cause an alarm signal because the KS test uses only the maximum separation in probability of cumulative probability distributions of concentration. The concentration range is not considered. Changes in the cumulative probability near the median concentration value in the data can occur as a consequence of small offsets that are not practically significant.

In the present work, data are corrected using eq. 6 if one or more alarms are triggered.  Whilst alarms are here based solely on the KS and MV tests applied without restrictions, this is a rule-based framework that is easily extendable to include other indications, for example knowledge that the proxy is reliable only for certain wind directions, or failure diagnostic signals derived from the sensor itself (Bart et al., 2014; Weissert et al., 2017). The management framework originally set out by Miskell et al.(Miskell et al., 2016) indeed used diagnostic signals derived directly from the sensor as well as the proxy comparisons. We did not use those signals in the present work, seeking only to evaluate the robustness of the proxy approach. The mean absolute bias (MAB) and pair-wise Pearson correlation coefficient ($R^2$) were used to quantify the accuracy and precision of the sensor data to the co-located regulatory analyzer data. All statistical analysis was in R (v 3.5.0) using the packages 'tidyverse' (Wickham, 2017), 'zoo' (Zeileis and Grothendieck, 2005), 'ggmap' (Kahle and Wickham, 2013), 'ggrepel' (Slowikowski, 2018) and 'lubridate' (Grolemond and Wickman, 2011).

## 3. Results

### 3.1 Overall performance of the low-cost sensor network

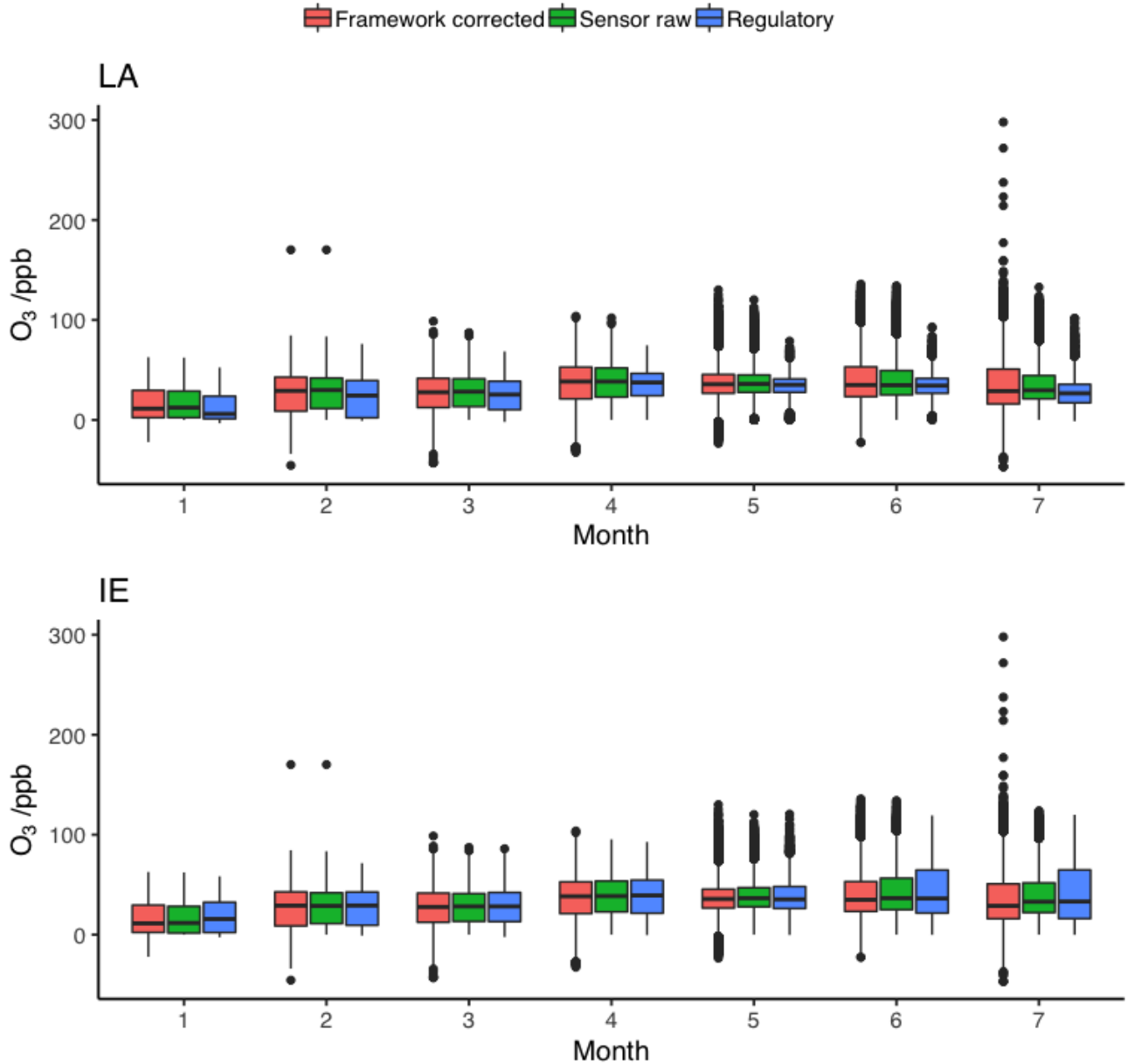

*Figure 3: Comparison of the hourly-averaged pooled data distributions from the regulatory station network (blue, right-hand boxes; n = 5 in each area) and with that of the raw data from the low-cost sensor network (green; centre boxes; n ~ 45 in IE and ~ 20 in LA as the exact numbers of operating sensors changed over time) and for the sensor network data corrected according to the management framework described in the text (red; left-hand boxes) for the two different local areas, by time since deployment. Month 1 is January 2018 – 7 is July 2018.*

Figure 3 compares the distribution of the entire dataset for the low-cost sensors with that for the regulatory instruments, over the full period of the study. Figure 3 also includes the sensor network data set corrected according to the management framework, as detailed later in the paper. Network median $O_3$ values were around 25 ppb in the IE network and around 30 ppb in the LA network. On average, the low-cost sensor network reliably represented the $O_3$ concentrations as reported by the regulatory network. The low-cost sensor networks had significantly larger numbers of measurement locations hence would be expected to capture greater variability, which might also be seasonally-dependent. The effect of the local variability is evident, in the wider spread of the corrected sensor network results, which is noticeable in month 7. The effect of drift of individual devices is seen in the narrower spread of outliers in comparison with the corrected network results.

### 3.2 Choice of proxy

In our previous work, we used land-use similarity as a criterion for choice of proxy. In the greater Los Angeles area, the land-use is very mixed, so this was not a criterion that could be applied unambiguously. Table 2 shows the dominant land-use characteristic of the various regulatory sites, where sensors were co-located. The annual average density of vehicle traffic (AADT) within 5 km is also listed. We used the regulatory network data to address the question: what constitutes a reliable proxy? We compared each site in the regulatory network to different choices for proxy: (a) the closest independent regulatory site; (b) the median of all the local network (LA or IE) measurements (low-cost sensor measurements – $n$ ~ 20 for LA and $n$ ~ 45 for IE - and $n$ = 5 regulatory measurements); or (c) an independent regulatory analyzer with a similar surrounding AADT. We used the median for speed and convenience of computation because using the pooled data lumped together was computationally very slow. The full network raw data median as a proxy was good in the earlier months but compromised by sensor drift in the later months of the study. In the Supplementary Information (SI), plots are given showing the correlation of the proxy-corrected data with the actual data for all three choices of proxy. Figure 4 gives a summary treating the regulatory analyzers as the test sites, comparing two different proxy choices and showing the fraction of the total time that each alarm was signaled, the MAB and the Pearson correlation coefficient, $R^2$, between the framework-corrected data and the actual site data – again, just for the regulatory station data.

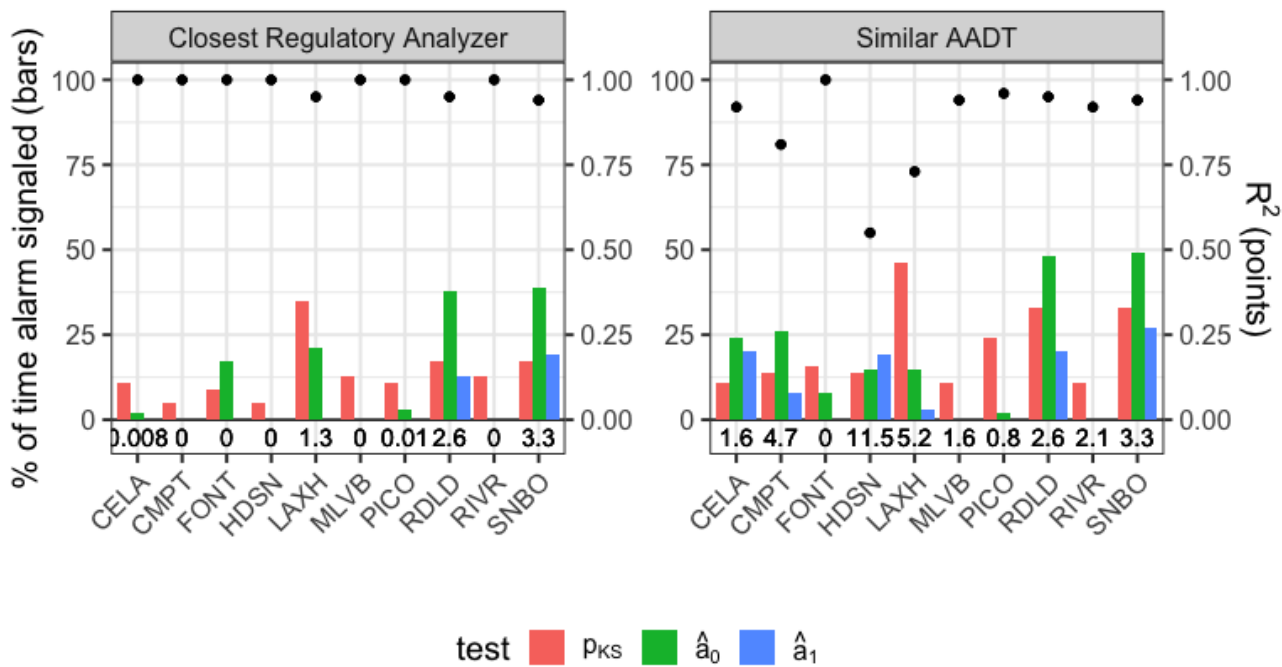

*Figure 4: Comparison of the different proxies (each panel: top) across sites using hourly-averaged regulatory data as the observations from the entire test period (seven months). Colored bars denote the percentage of total time that a test signaled an alarm during monitoring (left-axis), the black points denote the $R^2$ value (right-axis), and the written numbers near the site names denote the mean absolute bias values (ppb). The proxy corrections all used $t_d$ = 72 hr and $t_f$ = 120 hr. Site names are given and marked on Figure 1.*

If the proxy were perfect, the MAB would, of course, be zero and the Pearson correlation coefficient would be 1. The MAB shows the noise introduced as a consequence of proxy matching and the Pearson correlation coefficient shows the accuracy of the result. The proxy with the lowest proportion of alarm indications, the smallest MAB and the highest correlation of corrected to actual data for all the sites was the simplest: a reference station in closest proximity. One site, RDLD, had a much narrower distribution biased to higher values than other sites in the network. This site is known as one where $O_3$ concentrations may be different from elsewhere in the region (Epstein et al., 2017; Karamchandani et al., 2017). It is unusual in relation to others in the region because it is within a basin downwind of Los Angeles, that can lead to accumulation of $O_3$ (Neuman et al., 2012b), has a relatively low AADT so that the titration of $O_3$ by NO is less, and is set away from major roads. Its proxy pair based on proximity was the regulatory site SNBO, which although it had a similar AADT estimate, also had a higher density of roads within its vicinity compared to RDLD. The area around the regulatory site SNBO is also impacted by railyards and industry. However, even in this worst case the MAB introduced by the correction procedure was satisfactory: indeed, the correlation plots in the SI show that the errors were greatest at low $O_3$ concentration and that the high $O_3$ episodes were well captured by proxy matching. The choice of a proxy with a similar surrounding AADT, whilst satisfactory in some cases in others introduced significant errors (see also SI).

The question whether the proxy model is stable, and whether variations in the proxy model (eq. 2) can be distinguished from the effects of drift or interferences on the measurement model (eq 3) can to some extent be addressed by considering the time series of the MV slope parameter, $\hat{a}_1$. If the random error in the measurement model is small in comparison with that in the proxy model, then:

$$\hat{a}_1^2 \approx \frac{a_1^2}{b_1^2} + \frac{\text{var}\langle e_{j,t-t_d:t}\rangle}{b_1^2 \text{var}\langle Y_{j,t-t_d:t}\rangle} \qquad (6)$$

The time variation of $\hat{a}_1$ can be split into a long-term trend and short-term fluctuations about this trend. The long term trend can be attributed to drift in the sensor and possible seasonal variations in the coefficient $b_1$, and the short term fluctuations to short term variations in the assumed proxy correlation, reflected by variations in the parameters and in the error term, $e_j$. Evaluation of the reference station data against reference proxies can be used to distinguish these effects. Figure 5 shows the results. The proxy was for most sites stable and fluctuations died out on a timescale of less than 1 week. The timescales $t_d$ and $t_f$ used in the management framework are consistent with this. For the site at RDLD, for a period of approximately 1 month the trend moved outside the bounds set as appropriate for indicative monitoring, though not greatly so. A similar effect was observable at the site LAXH, where, associated with this trend there was a period where the short-term fluctuations were greater. Given this clear separation of short-term and long-term variations, the management framework was modified to use the long-term trend values of $\hat{a}_1$ and $\hat{a}_0$ for assessment and correction of the sensor results. The effect is to remove some of the noise attributable to proxy matching.

### 3.3 Management framework results for the low-cost sensors

Figure 6 shows the time variation of the MV slope, $\hat{a}_1$, for the sensors co-located at regulatory sites, using the closest other regulatory station as proxy. The long-term trend clearly picks out devices that drifted. The fluctuations about the trend are similar to those shown in figure 5.

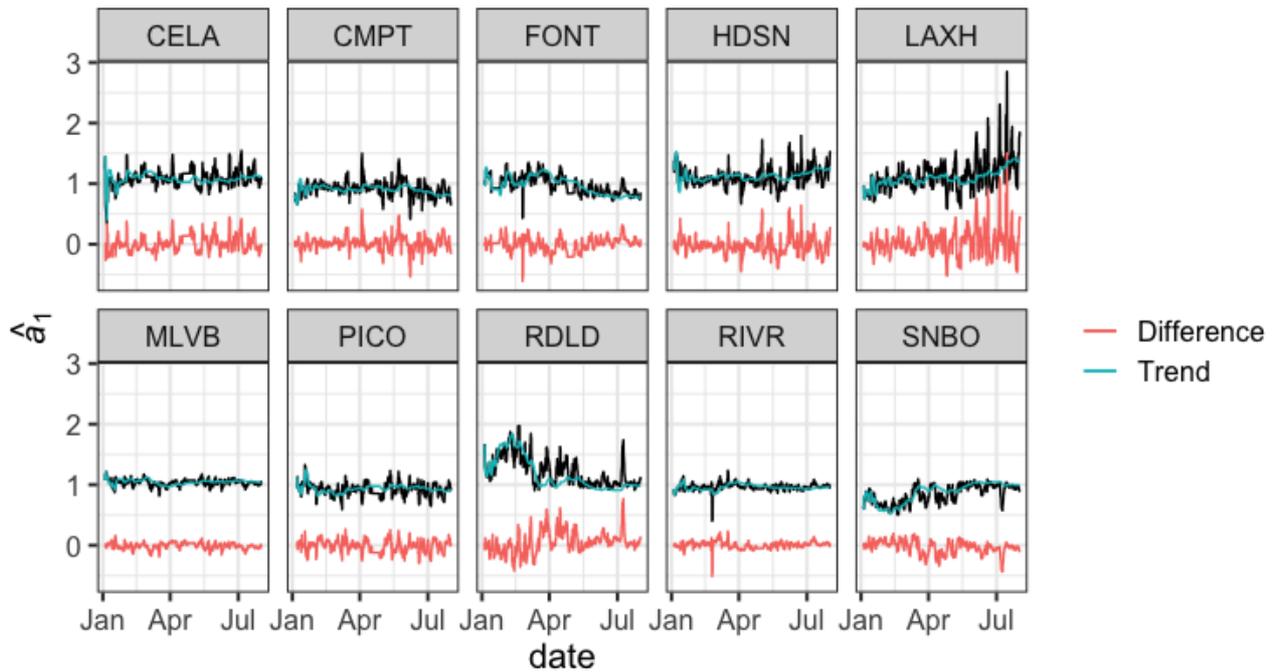

*Figure 5. Time series of the MV slope, $\hat{a}_1$, together with the long-term trend and the difference between the trend and the actual value, for the regulatory stations evaluated against the closest other regulatory station as proxy. The long-term trend value at time, t, is calculated as the least squares quadratic regression line from time t to time zero (the commencement of the test).*

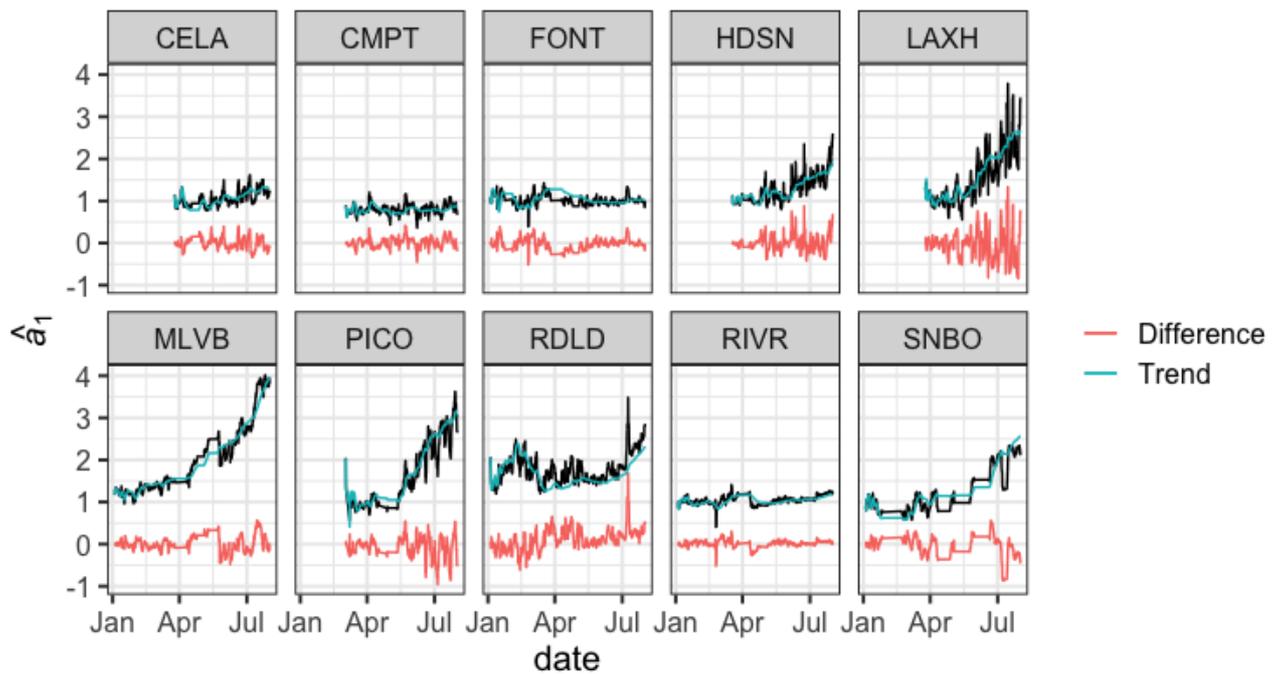

*Figure 6. Time series of the MV slope, $\hat{a}_1$, together with the long-term trend and the difference between the trend and the actual value, for the sensors co-located at regulatory stations evaluated*

*against the closest other regulatory station as proxy. The long-term trend value at time, t, is calculated as the least squares quadratic regression line from time t to time zero (the commencement of the test).*

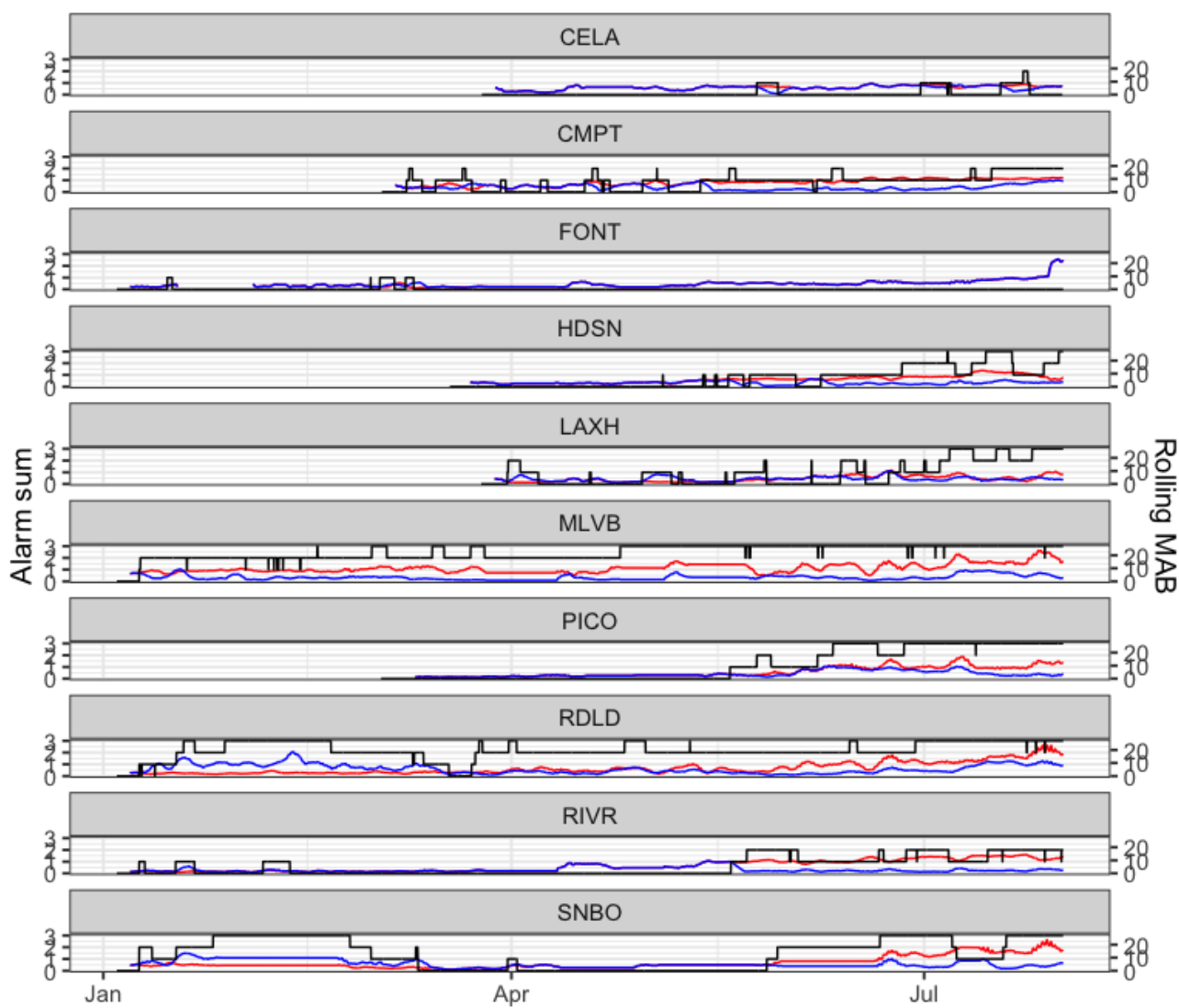

*Figure 7: Number of alarm signals generated by the low-cost sensor data in comparison with the proxy data (left-axis), and mean absolute bias (MAB) / ppb running over 72 hr of the low-cost sensor data with respect to the reference station at the same site (right-axis). The colored lines are the alarm sum (black), the uncorrected instrument MAB (red) and the framework-corrected MAB (blue). The framework correction uses the values of $\hat{a}_1$ and $\hat{a}_0$ derived from the quadratic long-term trend as shown on figure 6.*

Figure 7 shows for each individual low-cost sensor the evolution over time of the number of error signals generated by the proxy comparison and the MAB of both the uncorrected and framework-corrected data in relation to the regulatory analyzer with which the sensor was co-located. The sensor at MLVB generated an error signal immediately on deployment, implying a factory mis-calibration, that was trapped and corrected by the framework.

Figure 8 further compares the MAB variation with and without the framework correction. The closest proximity regulatory analyzer was chosen as the proxy. The framework-corrected MAB was within guidelines for indicative monitoring(U.S. Environmental Protection Agency, 2013) throughout the measurement campaign. Figure 9 shows the correlation plots of the framework-corrected low-cost sensor result against the co-located regulatory data, at the end of the study period.

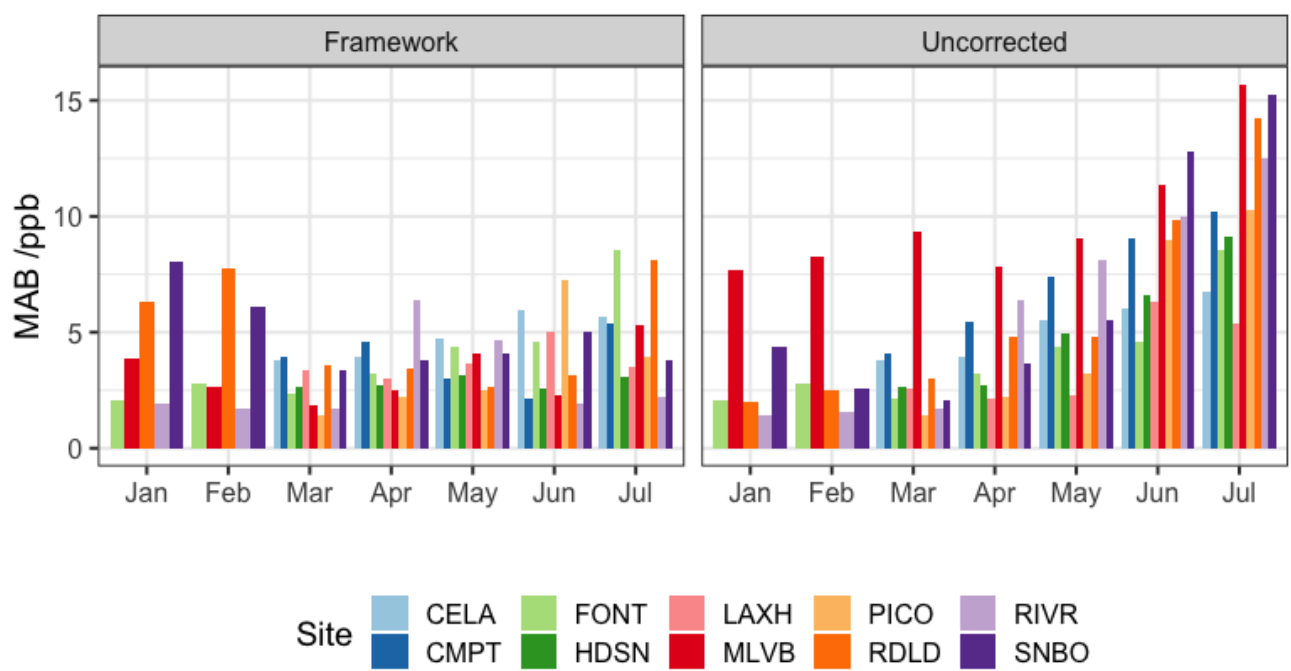

Figure 8: Monthly (1-7) averaged mean absolute bias (MAB) across the low-cost sensors compared with the regulatory station at the same site (left: framework-corrected data, right: uncorrected data). The framework correction uses the values of $\hat{a}_1$ and $\hat{a}_0$ derived from the quadratic long-term trend as shown on figure 6.

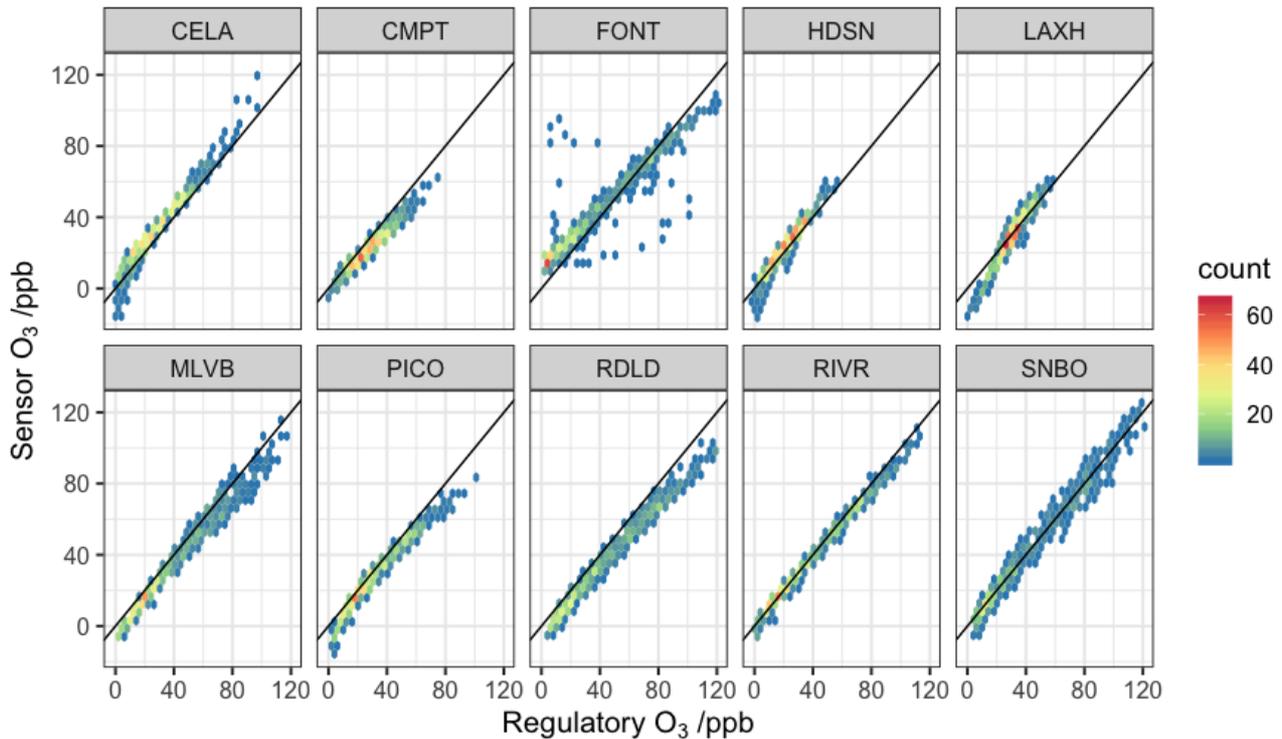

*Figure 9: Hex-bin scatterplots of the hourly-averaged framework-corrected data (n = 740) from the low-cost sensors against data from the regulatory instrument with which each one is co-located, for month 7 of deployment. The black line is the 1:1 line. The framework correction uses the values of $\hat{a}_1$ and $\hat{a}_0$ derived from the quadratic long-term trend as shown on figure 6. For the entire month 7 dataset, the root mean square deviation from the regulatory instruments was ±1.3 ppb.*

Some sensors had been in the field for seven months, others for four months (installation at month three). Many of the devices were stable for the full study period of seven months. In general, the framework detected the drifts, which inspection showed were caused by dirt depositing on the inlet filter, blocking it and consequently decreasing the airflow over the detection element (Williams et al., 2013). Figures 8 and 9 show that the management framework was successful in detecting and correcting the drift and keeping the MAB within the range 2-8 ppb over seven months: an improvement to the uncorrected data where MAB values showed a steady increase over time which ended with a range of 6-16 ppb. For the proxy pair RDLD-SNBO, the unusual data distribution at RDLD led to all three alarm signals registering almost immediately after the sensors were installed. Such an occurrence, in the absence of other information, would indicate either a sensor mis-calibration or that the proxy and test site did not satisfy the condition of similarity of data distribution. The result was overcorrection at low $O_3$ concentration, and an increase in the MAB, as noted above. However, the corrected low-cost sensor data captured the high concentrations reliably, including following a significant sensor drift, even though the proxy was not ideal. Full summary statistics are in the SI.

Control charts showing the variation of each of the test statistics for each of the low-cost sensors are given in the SI. Blockage of the inlet such that the sensor became essentially insensitive to $O_3$ was clearly signaled by $\hat{a}_0$ and $\hat{a}_1$ going to unacceptably large values and was also signaled by the sensor power consumption dropping significantly.

### 3.4 A mobile "buddy" for low-cost sensor checking

The unusual site RDLD brings into focus the question of how one can distinguish sensor mis-calibration from site-specific effects(Alavi-Shoshtari et al., 2018). One method is to use a mobile calibration device as a "buddy", that is calibrated at a regulatory site, moved to co-locate with the device to be checked, then moved back to the regulatory site for a second calibration validation. The low-cost sensors are easy to move and remount, so we tested this idea. Figure 10 shows results for three different sites, not at regulatory stations, where the sensors located there had data corrected using the management framework with the regulatory station in closest proximity as the proxy. The transfer "buddy" and the local low-cost sensor agree within ±10 ppb. The calibration of the transfer "buddy" was unaffected by the move. The result illustrates the feasibility of using the low-cost sensors as mobile devices to check calibration by "buddy" co-location. A map showing the instrument locations and movement is in the SI.

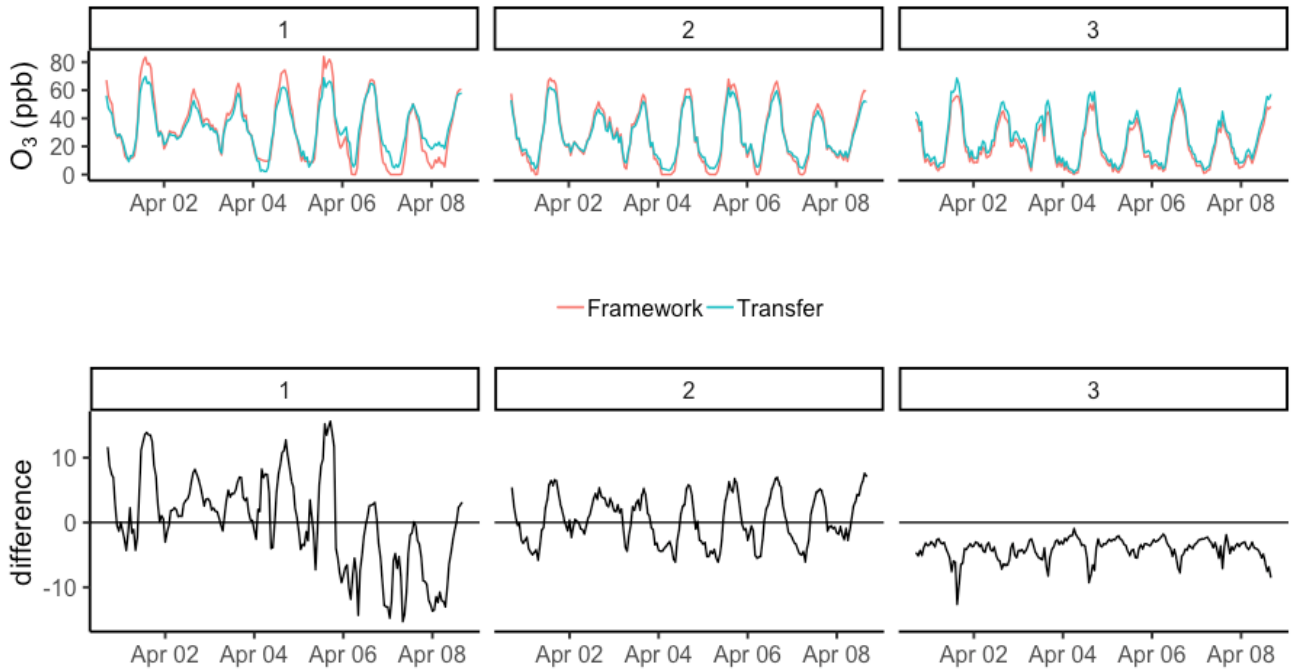

*Figure 10: Data validation by co-location of a calibrated "buddy". Low-cost sensors were first calibrated by co-location at a regulatory station, then moved to a site to be checked, then moved back to the regulatory station. The sensors being checked were managed using the closest proximity regulatory station as proxy. Top: ozone signal from the two sensors at the three sites. Bottom: difference / ppb between "buddy" and local sensor.*

### 3.5 Large local-scale spatial variations in ozone concentration revealed by the low-cost sensor network.

The purpose of the low-cost network has been stated as the extension of a regulatory network to capture neighborhood-scale variations. The method that we have described uses the regulatory network both to determine and validate the choice of proxy, and then to use the proxy distribution matching to check and re-calibrate if necessary the low-cost sensor network. Indeed, low-cost sensor network revealed significant ozone concentration variations that were not captured by the regulatory network, as illustrated in figure 11.

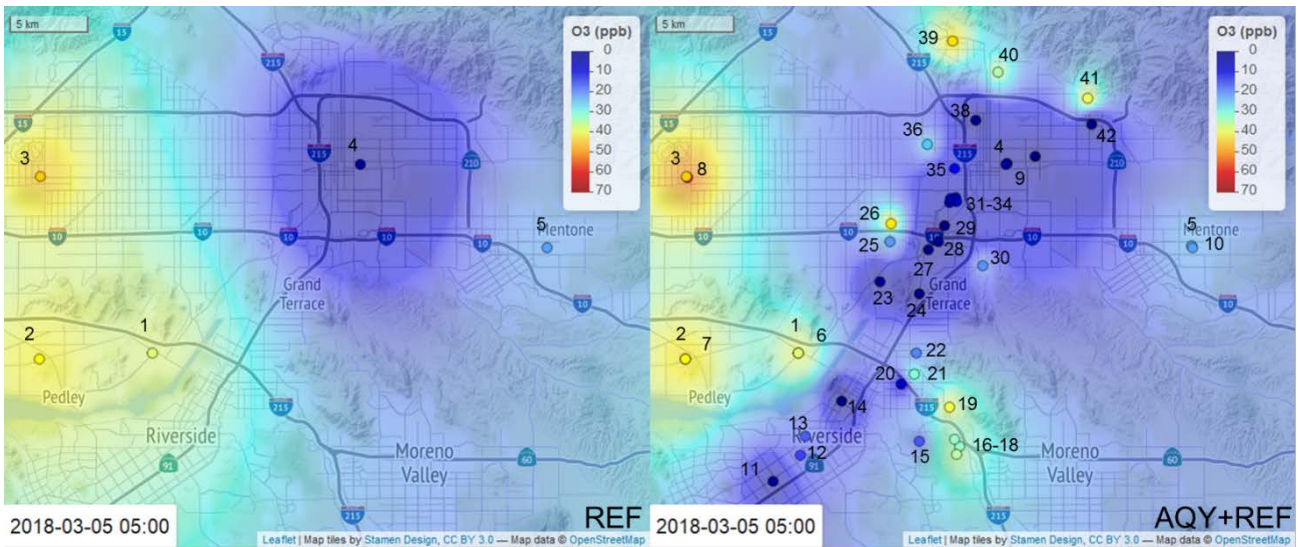

Figure 11. *Example of neighbourhood--scale variation in ozone concentration revealed by the low-cost sensor network. Left: reference network only; Right: reference network and low-cost sensor network. Interpolation by inverse-distance weighting (power, -2)*

The low-cost network reveals the ozone depletion in the valley along part of the highway network. It also shows the variability in this depletion and the high ozone concentrations a short distance from the highway network. These spatial variations are also strongly time-dependent, as shown by the time series for two sites in close proximity (41 and 42 on figure 11) given in figure 12. This figure shows also the traffic flow on highway 210 at the same time, the wind speed and direction averaged over the period shown and the detail of the location. A similar picture was seen for other sites such as 19 and 20.

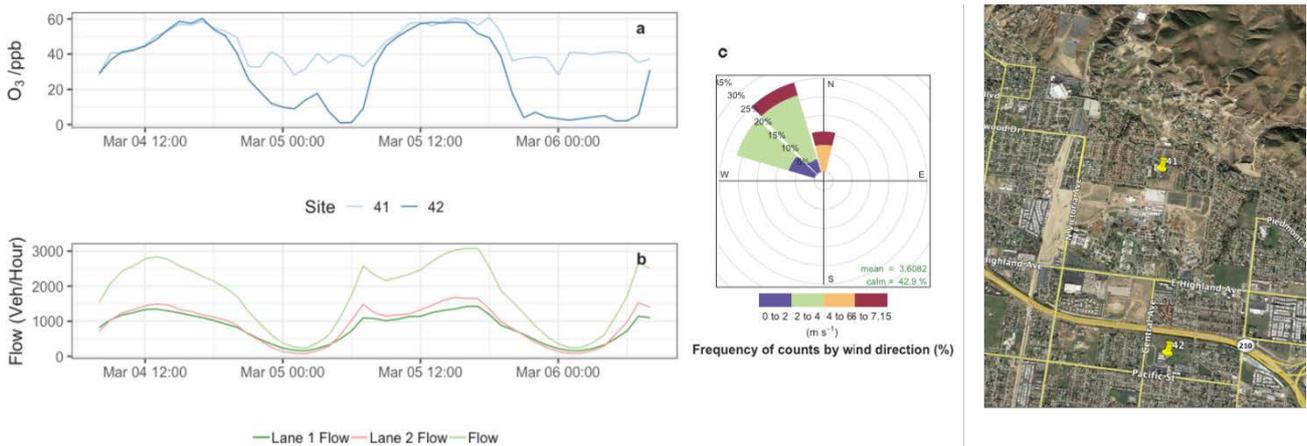

Figure 12: *Ozone variation at two locations in close proximity, with traffic flow on the nearby highway (210; source: http://pems.dot.ca.gov), wind speed and direction over the time shown, and aerial image of the area, showing the location of the two sites.*

The map can be interpreted as showing ozone titration by vehicle-emitted NO during the night, when the traffic flow remained high (~ 500 / hr).  During the day, when the traffic flow was extremely high, photochemical ozone production through photodissociation of nitrogen dioxide would cause elevated ozone levels near the highway (site 42 and other sites shown on figure 11). Although advection of the vehicle emissions from the highway towards site 41 would occur during the night, the ozone concentration at this site, though diminished, tended to remain high there throughout the night.  Ozone production, dispersion, long-range transport from the Pacific Ocean and transport to the surface from the free troposphere in California have been intensively studied (Bao et al., 2008; Neuman et al., 2012a; Ryerson et al., 2013).  The ground rises abruptly at the edge of the valley, close to site 41 and at other sites where similar effects are seen, such as 19.  One explanation, then, is a downslope drainage flow during the night transporting ozone from the upper troposphere to the valley floor at the edge of the valley (Bao et al., 2008) .  The effects are known to be complex(Caputi et al., 2018) and more work is required to clearly elucidate the underlying processes; however it is evident that dense networks of monitors, carefully managed to provide reliable data, can provide insight into the occurrence and causes of local air pollution hotspots. Such data also demonstrate that there is significant complexity in the spatio-temporal pattern of ozone concentrations, which would otherwise have gone undetected by the regulatory network.

## 4. Conclusion

We have demonstrated a hierarchical air quality measurement network, grounded in high-quality, compliant reference stations and extended to neighbourhood scale using low-cost sensors. We have validated a cost-effective approach to managing such air quality measurement networks and demonstrated that it delivers reliable results within an accepted specification for indicative measurement.  We have shown that a simple framework to both detect and correct observed drifts can be applied in a geographically complex area.  The key ideas are linearity of the sensor output and the use of a proxy measurement chosen to have a similar probability distribution averaged over diurnal variations.  The reference station network was used establish and validate the proxy choice.  A simple choice, the reference station in closest proximity, was satisfactory.  Even when the proxy had a rather different data distribution to the test site, the method was reliable in capturing the high ozone concentrations.  Provided the sensors satisfy the linearity condition, the framework provides reliable data from a low-cost sensor network.  The resultant map of ozone concentrations over the heavily trafficked area studied shows significant variations in both space and time, over small distance scales.

These significant small-scale variations were not captured by the reference network alone. Such a measurement network can now be used to answer granularity questions about urban air pollution, such as a more detailed examination of correlations between urban design and local-scale spatiotemporal air quality variation(Weissert et al., 2019).

## 5. Acknowledgements

This work was funded by the New Zealand Ministry for Business, Innovation and Employment, contract UOAX1413. This work was performed in collaboration with the Air Quality Sensor Performance Evaluation Center (AQ-SPEC) at the South Coast Air Quality Management District (SCAQMD). The authors would like to acknowledge the work of Mr. Berj Der Boghossian for his technical assistance with deploying AQY sensor nodes. The authors would like to acknowledge the work of the SCAQMD Atmospheric Measurements group of dedicated instrument specialists that operate, maintain, calibrate, and repair air monitoring instrumentation to produce regulatory-grade air monitoring data. DEW acknowledges the support of a fellowship at the Institute of Advanced Studies, Durham University, UK, and helpful discussions with Dr Jochen Einbeck and Dr Ostap Hryniv, of Durham University Mathematics Department.

## 6. Competing interests

KA and GSH are employees of Aeroqual Ltd, manufacturer of the sensor nodes used in the study. GSH and DEW are founders and shareholders in Aeroqual Ltd.

## 7. References

Ainslie, B., Steyn, D.G., 2007. Spatiotemporal trends in episodic ozone pollution in the Lower Fraser Valley, British Columbia, in relation to mesoscale atmospheric circulation patterns and emissions. Journal of Applied Meteorology and Climatology 46, 1631-1644.
Air Quality Sensor Performance Evaluation Center, 2018. Field evaluation Aeroqual AQY (v0.5). Retrieved from http://www.aqmd.gov/docs/default-source/aq-spec/field-evaluations/aeroqual-aqy-v0-5---field-evaluation.pdf?sfvrsn=4
Alavi-Shoshtari, M., Salmond, J.A., Giurcaneanu, C.D., Miskell, G., Weissert, L., Williams, D.E., 2018. Automated data scanning for dense networks of low-cost air quality instruments: Detection and differentiation of instrumental error and local to regional scale environmental abnormalities. Environmental Modelling & Software 101, 34-50.
Alavi-Shoshtari, M., Williams, D.E., Salmond, J.A., Kaipio, J.P., 2013. Detection of malfunctions in sensor networks. Environmetrics 24, 227-236.
Aliwell, S.R., Halsall, J.F., Pratt, K.F.E., O'Sullivan, J., Jones, R.L., Cox, R.A., Utembe, S.R., Hansford, G.M., Williams, D.E., 2001. Ozone sensors based on $WO_3$: a model for sensor drift and a measurement correction method. Measurement Science and Technology 12, 684-690.

Bao, J., Michelson, S., Persson, P., Djalalova, I., Wilczak, J., 2008. Observed and WRF-simulated low-level winds in a high-ozone episode during the Central California Ozone Study. Journal of Applied Meteorology and Climatology 47, 2372-2394.

Bart, M., Williams, D.E., Ainslie, B., McKendry, I., Salmond, J., Grange, S.K., Alavi-Shoshtari, M., Steyn, D., Henshaw, G.S., 2014. High Density Ozone Monitoring Using Gas Sensitive Semi-Conductor Sensors in the Lower Fraser Valley, British Columbia. Environmental Science & Technology 48, 3970-3977.

Bermudez, M., Fine, P., 2010. South Coast Air Quality Management District annual air quality monitoring network plan. Retrieved from https://www3.epa.gov/ttnamti1/files/networkplans/CASCAQMDPlan2010.pdf

British Columbia, 2016. George Massey tunnel replacement project: Traffic forecasting summary. Retrieved from https://engage.gov.bc.ca/app/uploads/sites/52/2017/02/GMT-2016-08-00-Traffic-Forecasting-Summary.pdf

Caputi, D.J., Faloona, I., Trousdell, J., Smoot, J., Falk, N., Conley, S., 2018. Residual Layer Ozone, Mixing, and the Nocturnal Jet in California's San Joaquin Valley. Atmos. Chem. Phys. Discuss., https://doi.org/10.5194/acp-2018-854  Manuscript under review for journal Atmos. Chem. Phys. .

Cavellin, L.D., Weichenthal, S., Tack, R., Ragettli, M.S., Smargiassi, A., Hatzopoulou, M., 2016. Investigating the Use Of Portable Air Pollution Sensors to Capture the Spatial Variability Of Traffic-Related Air Pollution. Environmental Science & Technology 50, 313-320.

Chameides, W.L., Fehsenfeld, F., Rodgers, M.O., Cardelino, C., Martinez, J., Parrish, D., Lonneman, W., Lawson, D.R., Rasmussen, R.A., Zimmerman, P., Greenberg, J., Middleton, P., Wang, T., 1992. Ozone precursor relationships in the ambient atmosphere. Journal of Geophysical Research-Atmospheres 97, 6037-6055.

Epstein, S.A., Lee, S.-M., Katzenstein, A.S., Carreras-Sospedra, M., Zhang, X., Farina, S.C., Vahmani, P., Fine, P.M., Ban-Weiss, G., 2017. Air-quality implications of widespread adoption of cool roofs on ozone and particulate matter in southern California. Proceedings of the National Academy of Sciences of the United States of America 114, 8991-8996.

Grolemond, G., Wickman, H., 2011. Dates and Times Made Easy with lubridate. Journal of Statistical Software 40, 1-25.

Hansford, G.M., Freshwater, R.A., Bosch, R.A., Cox, R.A., Jones, R.L., Pratt, K.F.E., Williams, D.E., 2005. A low cost instrument based on a solid state sensor for balloon-borne atmospheric $O_3$ profile sounding. Journal of Environmental Monitoring 7, 158-162.

Kahle, D., Wickham, H., 2013. ggmap: Spatial Visualization with ggplot2. R Journal 5, 144-161.

Karamchandani, P., Morris, R., Wentland, A., Shah, T., Reid, S., Lester, J., 2017. Dynamic Evaluation of Photochemical Grid Model Response to Emission Changes in the South Coast Air Basin in California. Atmosphere 8, 145.

Lin, C., Masey, N., Wu, H., Jackson, M., Carruthers, D.J., Reis, S., Doherty, R.M., Beverland, I.J., Heal, M.R., 2017. Practical Field Calibration of Portable Monitors for Mobile Measurements of Multiple Air Pollutants. Atmosphere 8, 231.

Lu, R., Turco, R.P., 1995. Air Pollutant Transport in a Coastal Environment .II. 3-Dimensional Simulations Over Los-Angeles Basin. Atmospheric Environment 29, 1499-1518.

Miskell, G., Salmond, J., Alavi-Shoshtari, M., Bart, M., Ainslie, B., Grange, S., McKendry, I.G., Henshaw, G.S., Williams, D.E., 2016. Data Verification Tools for Minimizing Management Costs of Dense Air-Quality Monitoring Networks. Environmental Science & Technology 50, 835-846.

Miskell, G., Salmond, J.A., Williams, D.E., 2018. Solution to the Problem of Calibration of Low-Cost Air Quality Measurement Sensors in Networks. ACS Sensors 3, 832-843.

Neuman, J.A., Trainer, M., Aikin, K.C., Angevine, W.M., Brioude, J., Brown, S.S., de Gouw, J.A., Dube, W.P., Flynn, J.H., Graus, M., Holloway, J.S., Lefer, B.L., Nedelec, P., Nowak, J.B., Parrish, D.D., Pollack, I.B., Roberts, J.M., Ryerson, T.B., Smit, H., Thouret, V., Wagner, N.L., 2012a. Observations of ozone transport from the free troposphere to the Los Angeles basin. Journal of Geophysical Research-Atmospheres 117, 15.

Neuman, J.A., Trainer, M., Aikin, K.C., Angevine, W.M., Brioude, J., Brown, S.S., de Gouw, J.A., Dube, W.P., Flynn, J.H., Graus, M., Holloway, J.S., Lefer, B.L., Nedelec, P., Nowak, J.B., Parrish, D.D., Pollack, I.B., Roberts, J.M., Ryerson, T.B., Smit, H., Thouret, V., Wagner, N.L., 2012b.

Observations of ozone transport from the free troposphere to the Los Angeles basin. Journal of Geophysical Research-Atmospheres 117, D00V09.

Pattinson, W., Kingham, S., Longley, I., Salmond, J., 2017. Potential pollution exposure reductions from small-distance bicycle lane separations. Journal of Transport & Health 4, 40-52.

Ryerson, T.B., Andrews, A.E., Angevine, W.M., Bates, T.S., Brock, C.A., Cairns, B., Cohen, R.C., Cooper, O.R., de Gouw, J.A., Fehsenfeld, F.C., Ferrare, R.A., Fischer, M.L., Flagan, R.C., Goldstein, A.H., Hair, J.W., Hardesty, R.M., Hostetler, C.A., Jimenez, J.L., Langford, A.O., McCauley, E., McKeen, S.A., Molina, L.T., Nenes, A., Oltmans, S.J., Parrish, D.D., Pederson, J.R., Pierce, R.B., Prather, K., Quinn, P.K., Seinfeld, J.H., Senff, C.J., Sorooshian, A., Stutz, J., Surratt, J.D., Trainer, M., Volkamer, R., Williams, E.J., Wofsy, S.C., 2013. The 2010 California Research at the Nexus of Air Quality and Climate Change (CalNex) field study. Journal of Geophysical Research: Atmospheres 118, 5830-5866.

Sadighi, K., Coffey, E., Polidori, A., Feenstra, B., Lv, Q., Henze, D.K., Hannigan, M., 2018. Intra-urban spatial variability of surface ozone in Riverside, CA: viability and validation of low-cost sensors. Atmospheric Measurement Techniques 11, 1777-1792.

Salmond, J., Sabel, C.E., Vardoulakis, S., 2018. Towards the Integrated Study of Urban Climate, Air Pollution, and Public Health. Climate 6.

Slowikowski, K., 2018. Ggrepel: automatically position non-overlapping text labels with 'ggplot2'. R package version 0.8.0.

South Coast Air Quality Management District, 2016. 2016 air quality data table. Retrieved from https://www.aqmd.gov/home/air-quality/air-quality-data-studies/historical-data-by-year

Statistics Canada, 2016. Population and dwelling counts, for Canada, provinces and territories, and census divisions, 2016 and 2011 censuses. Retrieved from http://www12.statcan.gc.ca/census-recensement/2016/dp-pd/hlt-fst/pd-pl/Table.cfm?Lang=Eng&T=701&SR=1&S=3&O=D&RPP=9999&PR=59

U.S. Census Bureau, 2010. Profile of general population and housing characteristics.

U.S. Department of Transportation, 2016. 2016 traffic volumes on California state highways. Retrieved from http://www.dot.ca.gov/trafficops/census/docs/2016_aadt_volumes.pdf

U.S. Environmental Protection Agency, 2006. Air quality criteria for ozone and related photochemical oxidants, EPA600/R-05/004aF, Research Triangle Park, NC.

U.S. Environmental Protection Agency, 2013. QA Handbook for air pollution measurement systems: Volume 2, ambient air quality monitoring program; report no EPA454/B-13/003, Research Traingle Park, NC.

Utembe, S.R., Hansford, G.M., Sanderson, M.G., Freshwater, R.A., Pratt, K.F.E., Williams, D.E., Cox, R.A., Jones, R.L., 2006. An ozone monitoring instrument based on the tungsten trioxide ($WO_3$) semiconductor. Sensors and Actuators B-Chemical 114, 507-512.

Weissert, L.F., Alberti, K., Miskell, G., Pattinson, W., Salmond, J.A., Henshaw, G., Williams, D.E., 2019. Low-cost sensors and microscale land use regression: data fusion to resolve air quality variations with high spatial and temporal resolution. submitted.

Weissert, L.F., Salmond, J.A., Miskell, G., Alavi-Shoshtari, M., Grange, S.K., Henshaw, G.S., Williams, D.E., 2017. Use of a dense monitoring network of low-cost instruments to observe local changes in the diurnal ozone cycles as marine air passes over a geographically isolated urban centre. Science of the Total Environment 575, 67-78.

Wickham, H., 2017. Tidyverse: easily install and load the 'tidyverse'. R package version 1.2.1.

Williams, D.E., Aliwell, S.R., Pratt, K.F.E., Caruana, D.J., Jones, R.L., Cox, R.A., Hansford, G.M., Halsall, J., 2002. Modelling the response of a tungsten oxide semiconductor as a gas sensor for the measurement of ozone. Measurement Science and Technology 13, 923-931.

Williams, D.E., Henshaw, G.S., Bart, M., Laing, G., Wagner, J., Naisbitt, S., Salmond, J.A., 2013. Validation of low-cost ozone measurement instruments suitable for use in an air-quality monitoring network. Measurement Science and Technology 24, 065803.

Zeileis, A., Grothendieck, G., 2005. zoo: S3 infrastructure for regular and irregular time series. Journal of Statistical Software 14.

Supporting Information

*Reliable data from low cost ozone sensors in a hierarchical network*

Georgia Miskell[1], Kyle Alberti[2], Brandon Feenstra[4], Geoff S Henshaw[2], Vasileios Papapostolou[4], Hamesh Patel[2], Andrea Polidori[4], Jennifer A Salmond[3], Lena Weissert[1,3], David E Williams[1,*]

1. School of Chemical Sciences and MacDiarmid Institute for Advanced Materials and Nanotechnology, University of Auckland, Private Bag 92019, Auckland 1142, New Zealand
2. Aeroqual Ltd, 460 Rosebank Road, Avondale, Auckland 1026, New Zealand
3. School of Environment, University of Auckland, Private Bag 92019, Auckland 1142, New Zealand
4. South Coast Air Quality Management District, 21865 Copley Drive, Diamond Bar, CA 91765, USA

1. The KS test
2. Regulatory observations compared with three different proxies
3. Co-located sensor control charts using the closest other regulatory station as the proxy
4. Summary results for the application of the framework on the sensor data, using closest other regulatory station as proxy.
5. Map showing the movement of "buddy" sensors from regulatory sites to test sites

1. The KS test

The Empirical Cumulative Distribution (ECD) for a sample with $t_w$ observations and variable *X* has the cumulative probability:

$$\mathbb{P}(x > X) = \hat{F}_{t_w}(x) = \frac{1}{t_w + 1} \sum_{i=1}^{t_w} \begin{cases} 1, & x > x_i \\ 0, & x \leq x_i \end{cases}$$

Assumptions are $x_i$ are *iid* random variables with a common underlying ECD, *F(X)*. The KS statistic uses the ECD over a specified time window, $t_w$. The test hypothesizes the two samples will have similar distributions by comparing the maximum absolute distance (supremum function, sup) between the cumulative probability curves on the *y*-axis for the same *x*-axis values. The KS statistic, *D,* for two samples with *m, $t_w$* observations and common variable *X* is:

$$D_{m,t_w} = \frac{sup}{x} |\hat{F}_m(x) - \hat{F}_{t_w}(x)|$$

KS statistics are translatable into *p*-values ($p_{KS}$) based on sample size and probability thresholds, with the null hypothesis, $H_0$, being the two samples could come from the same distribution. Assumptions are the two samples are independent, observations are independent and both come from continuous distributions.

2. Regulatory observations compared with three different proxies

Three proxies were tested against the regulatory data to examine their suitability as a proxy. The three proxies tested were: nearest regulatory station data not co-located with the instrument data under test, local network median (including regulatory and instrument data), and regulatory station with the most similar AADT within 5 km to the test location.

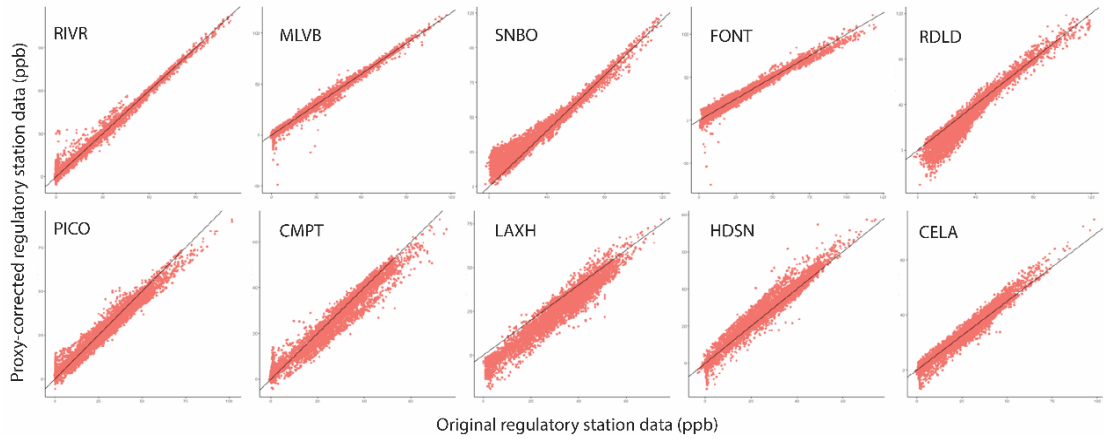

Nearest Regulatory Station as Proxy

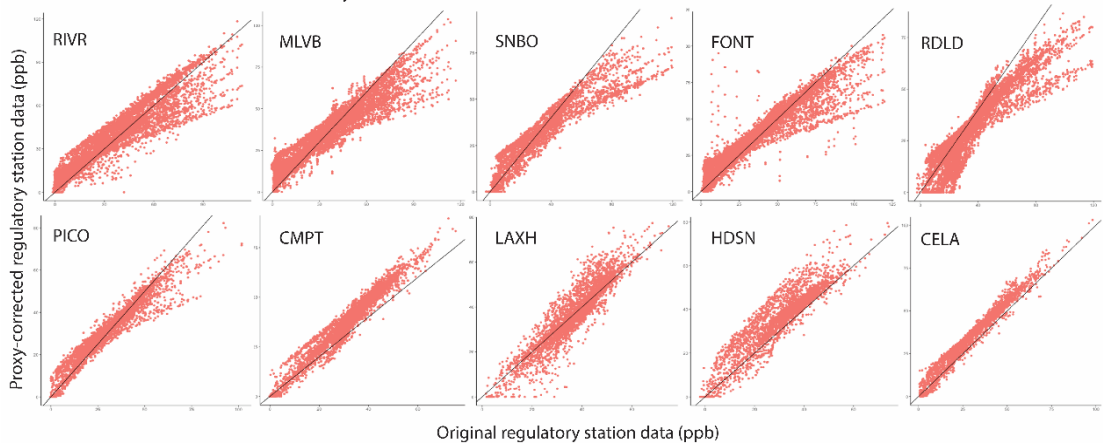

Local Network Median as Proxy

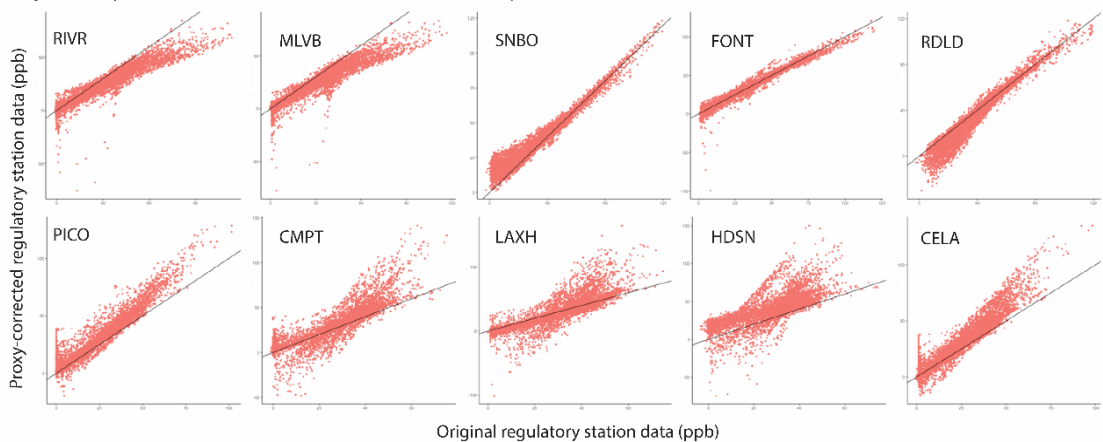

Regulatory Station with Similar AADT as Proxy

3. Co-located sensor control charts using the closest other regulatory station as the proxy

The control charts (four plots) for the ten co-located sensors. The four plots show the KS-test p-value ($p_{KS}$) and the MV-test intercept ($\hat{a}_0$) and slope ($\hat{a}_1$), with dashed lines representing the defined thresholds and the large black points where the test result is outside of the threshold for the defined length, $t_t$. The bottom plot (in black) shows the sum of the black points – the test alarms - over time. Where this value is two or above, the correction algorithm from the MV test is applied to the instrument data to derive the framework-corrected result.

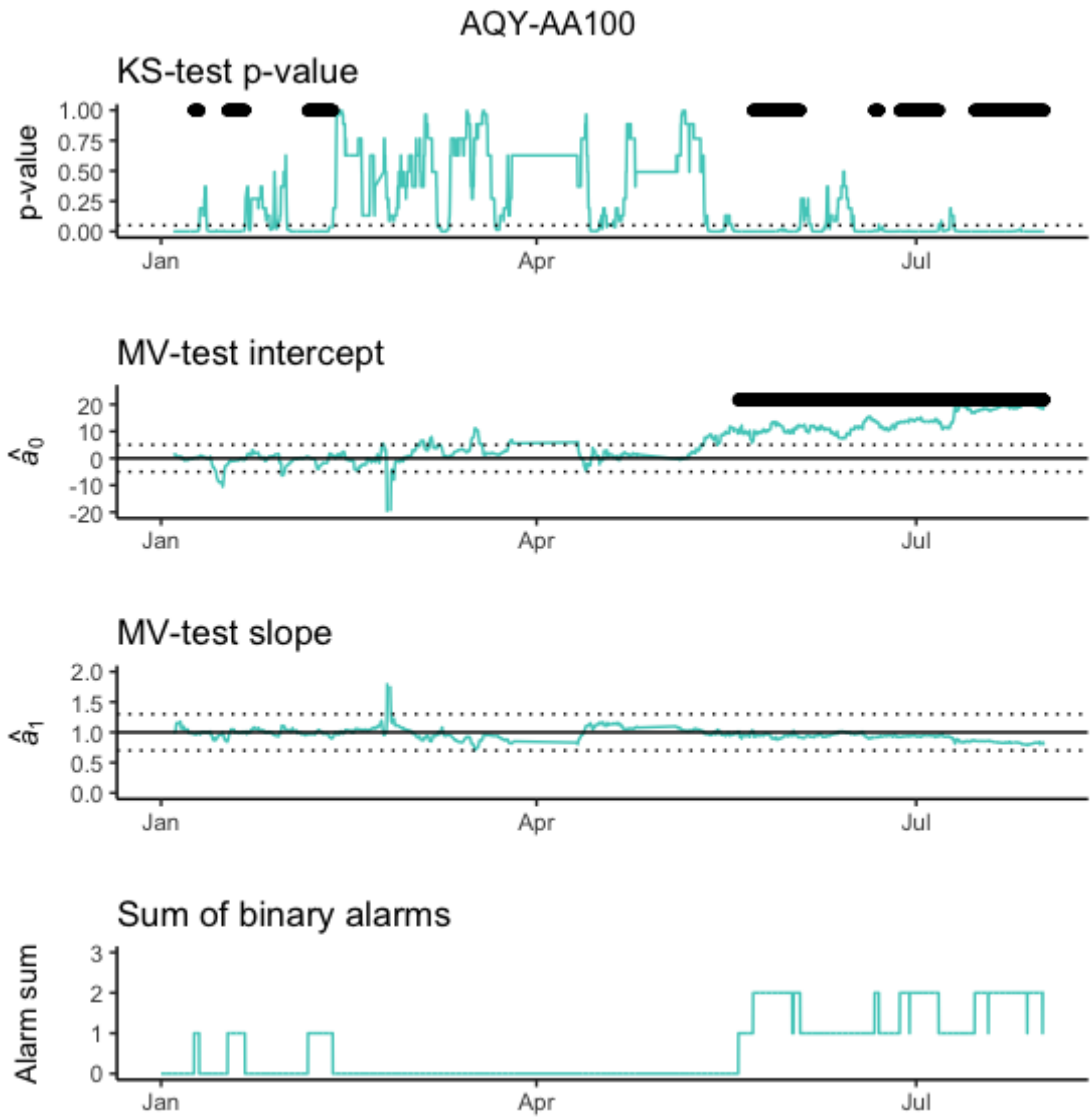

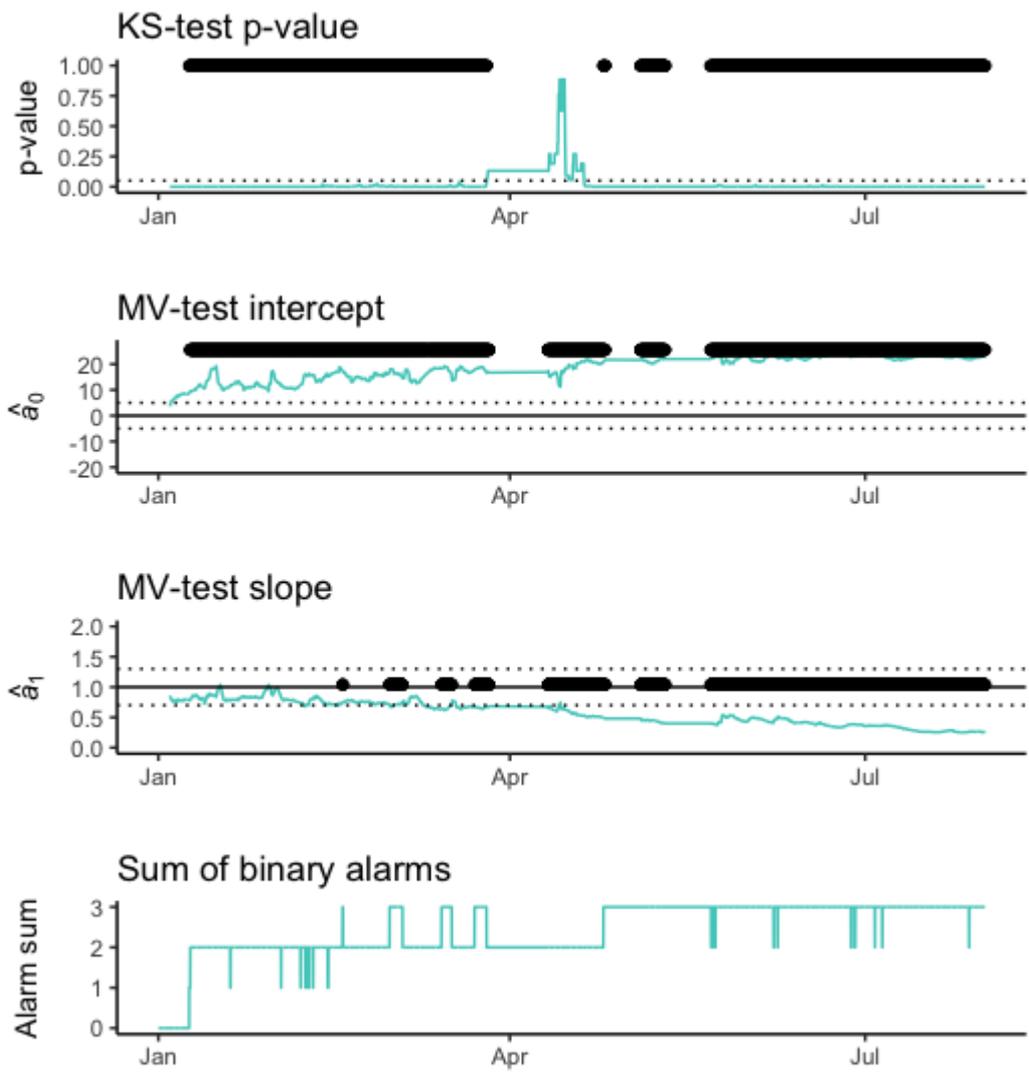

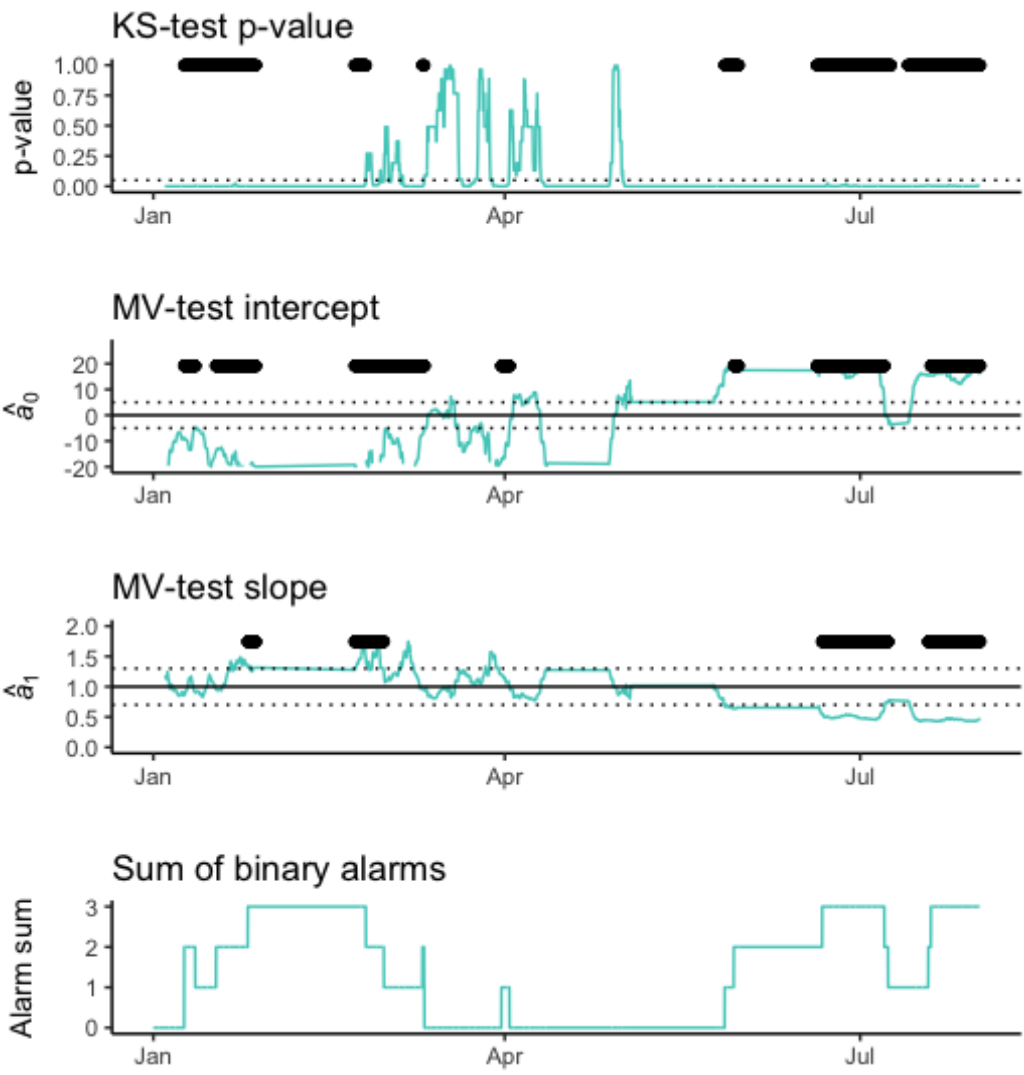

AQY-AA103

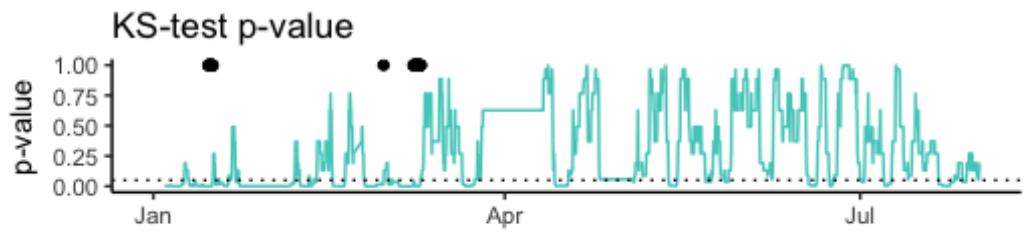

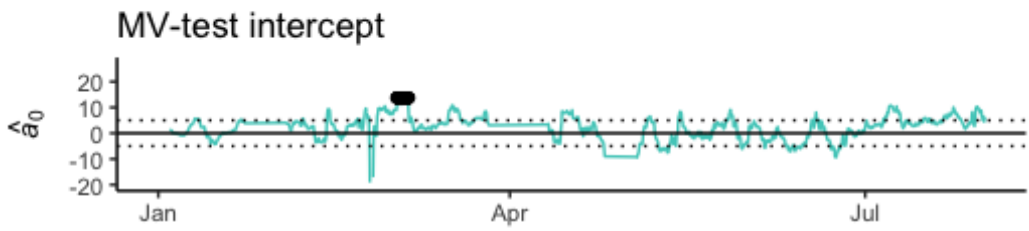

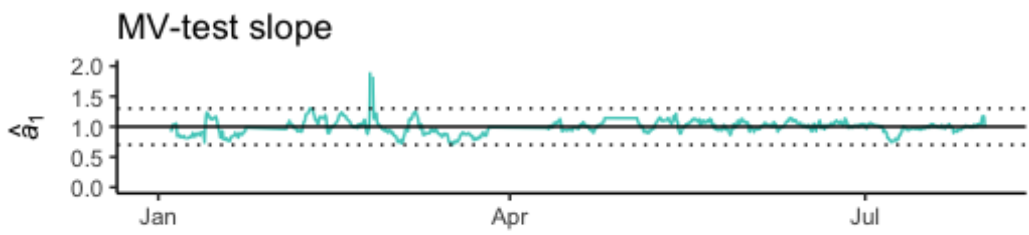

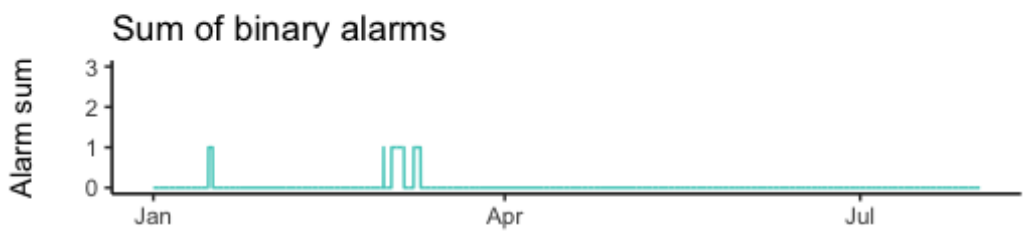

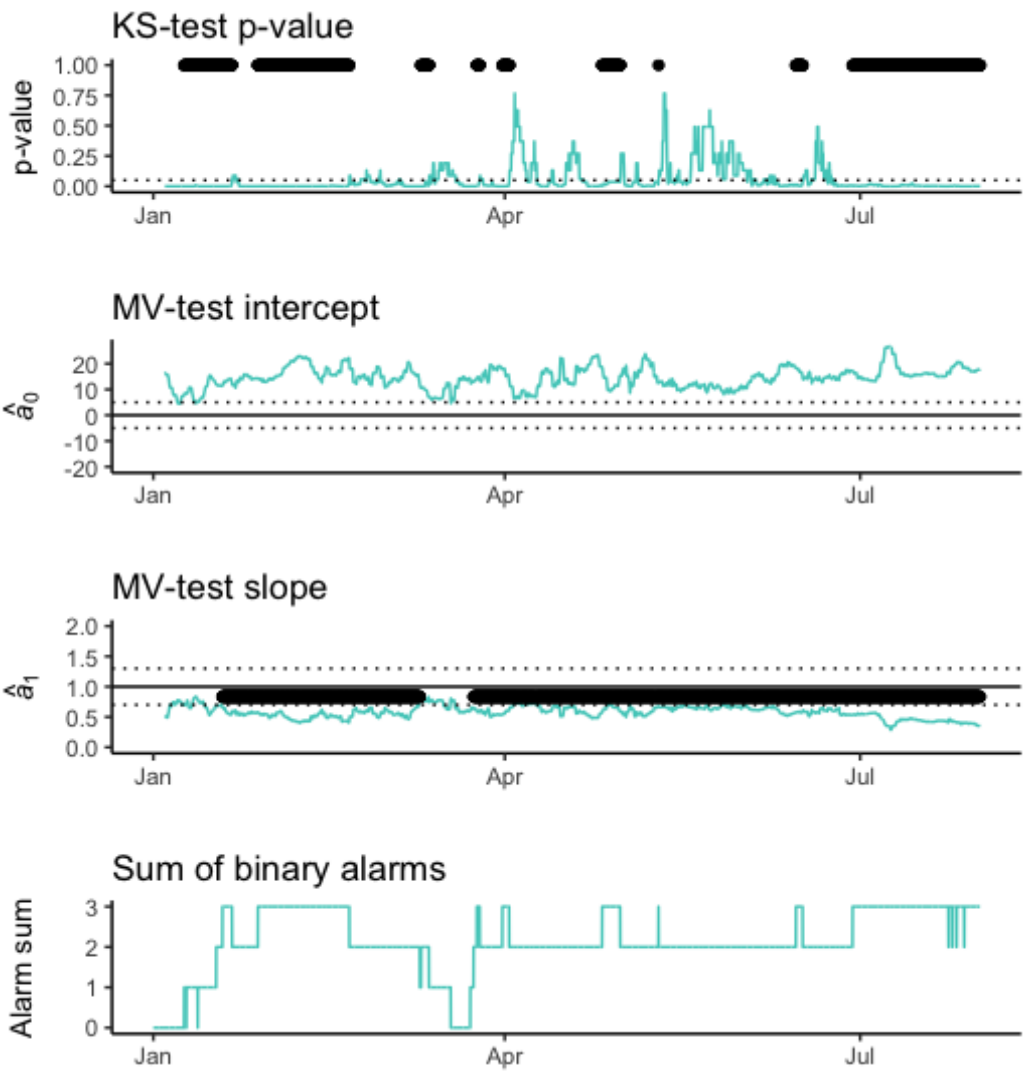

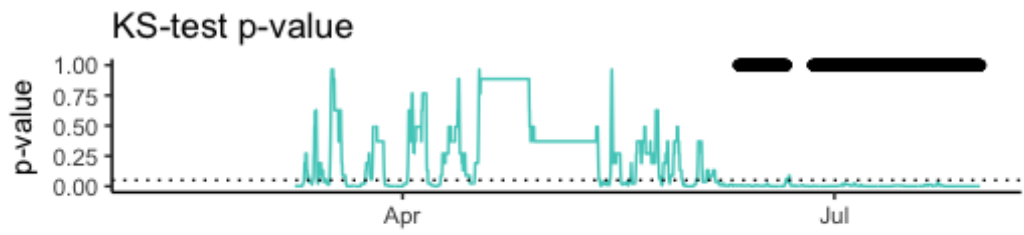
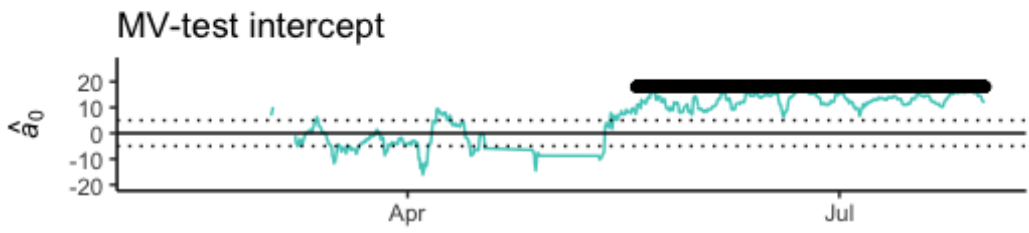
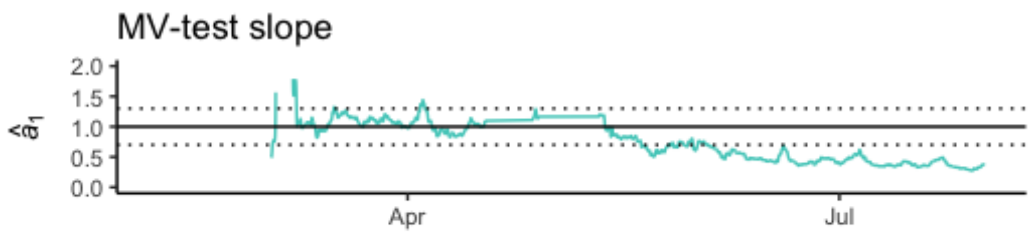
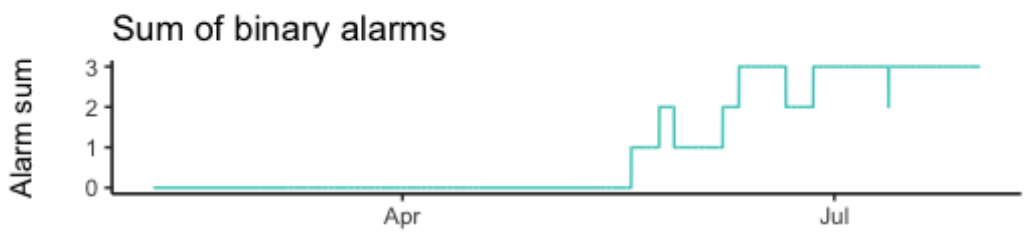

## AQY-AA166

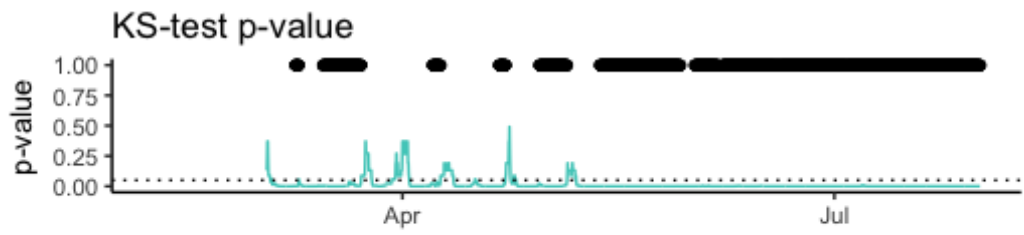

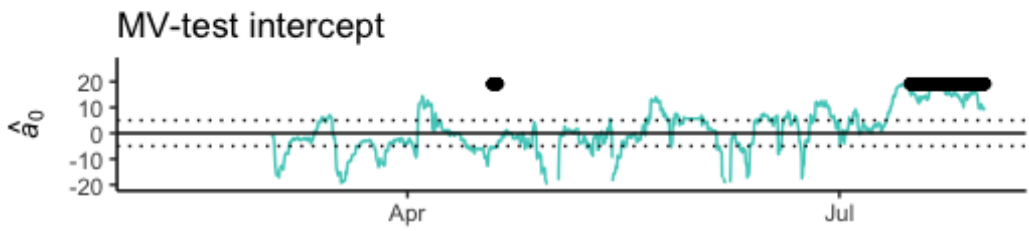

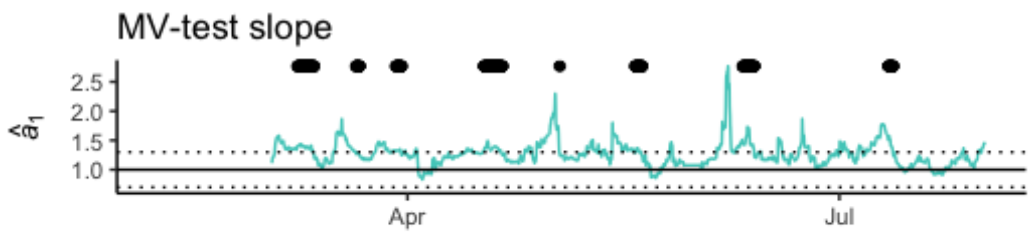

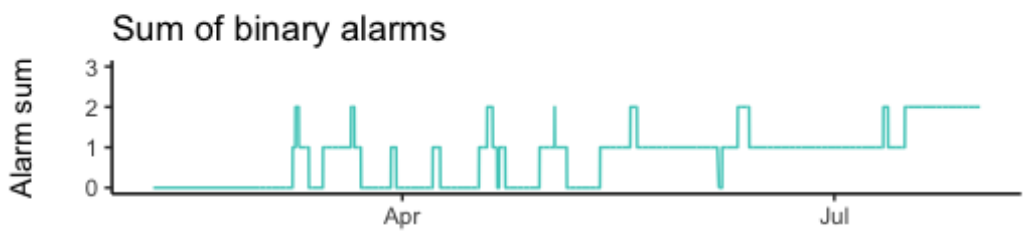

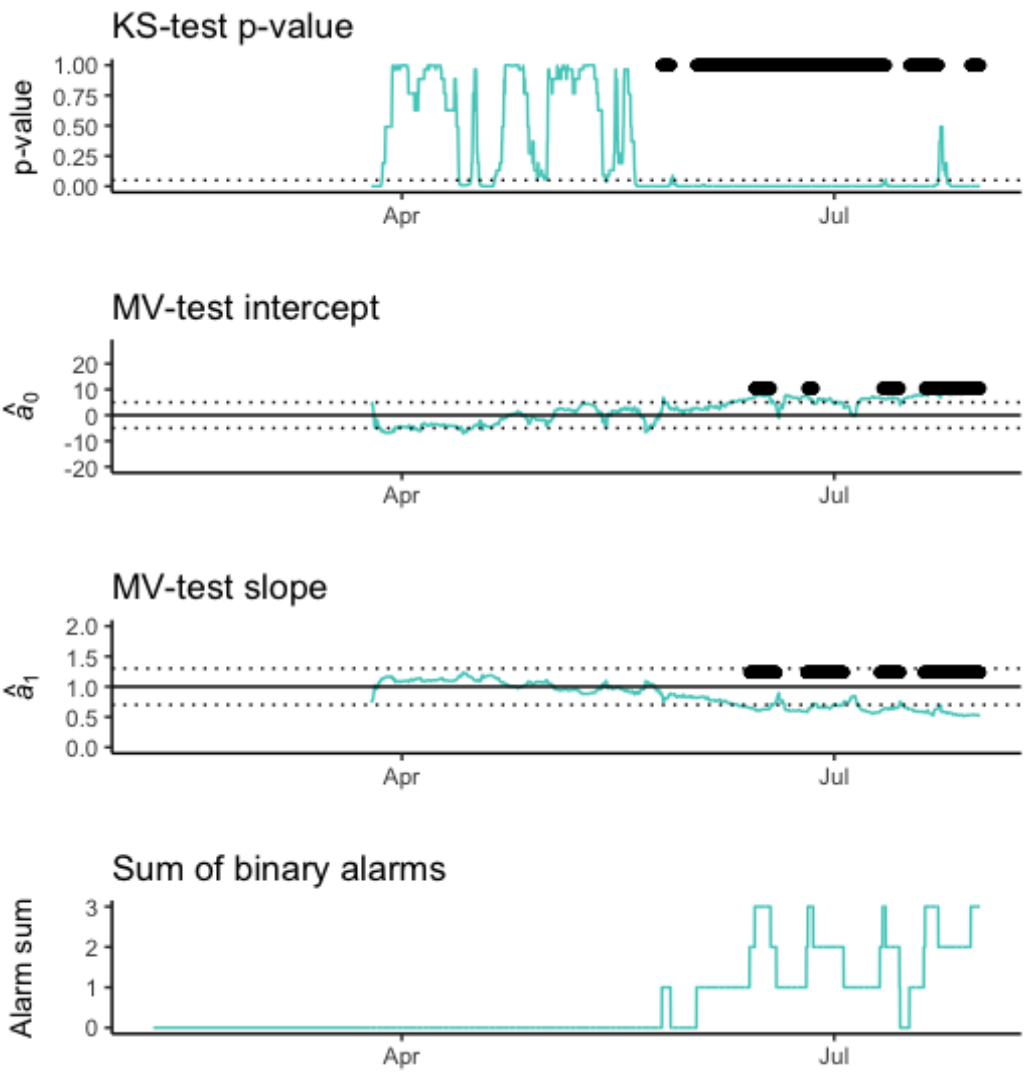

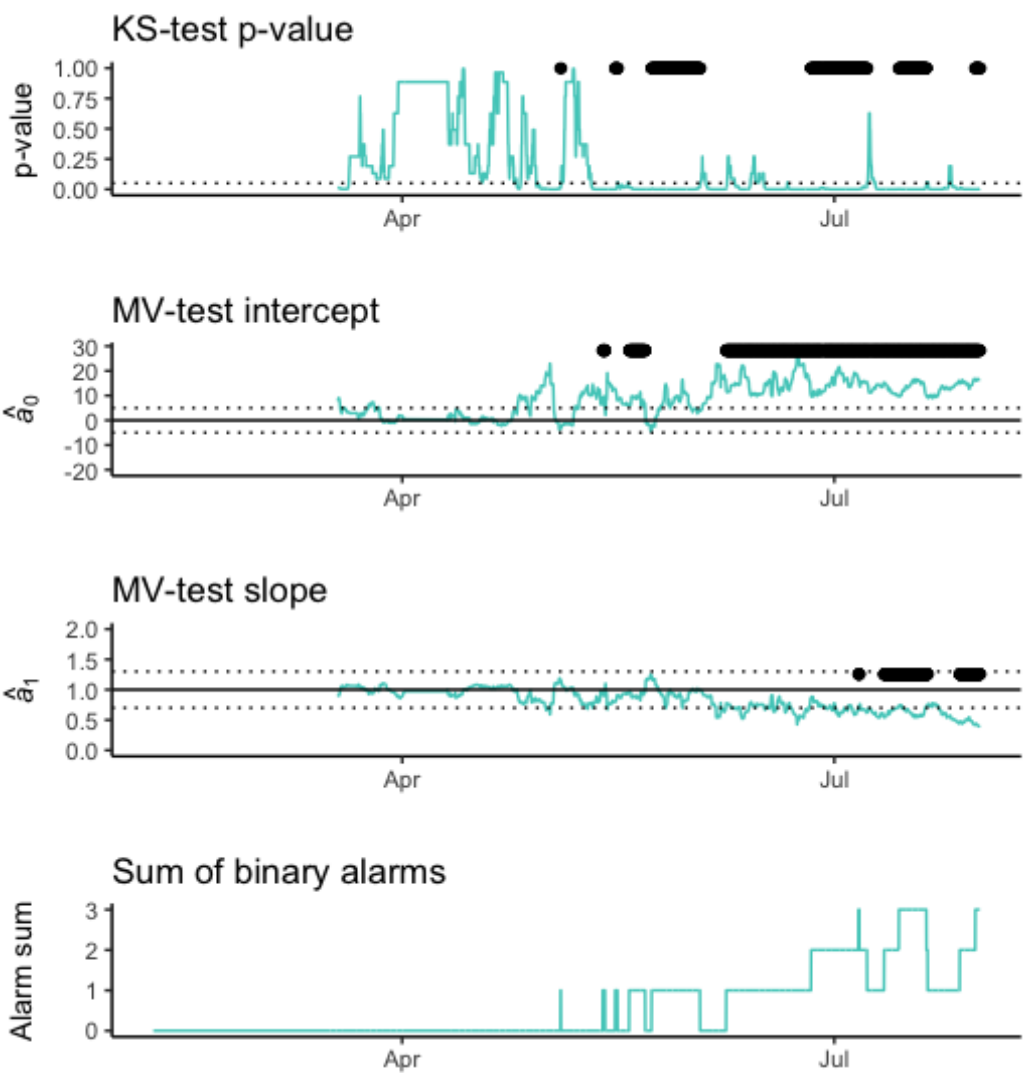

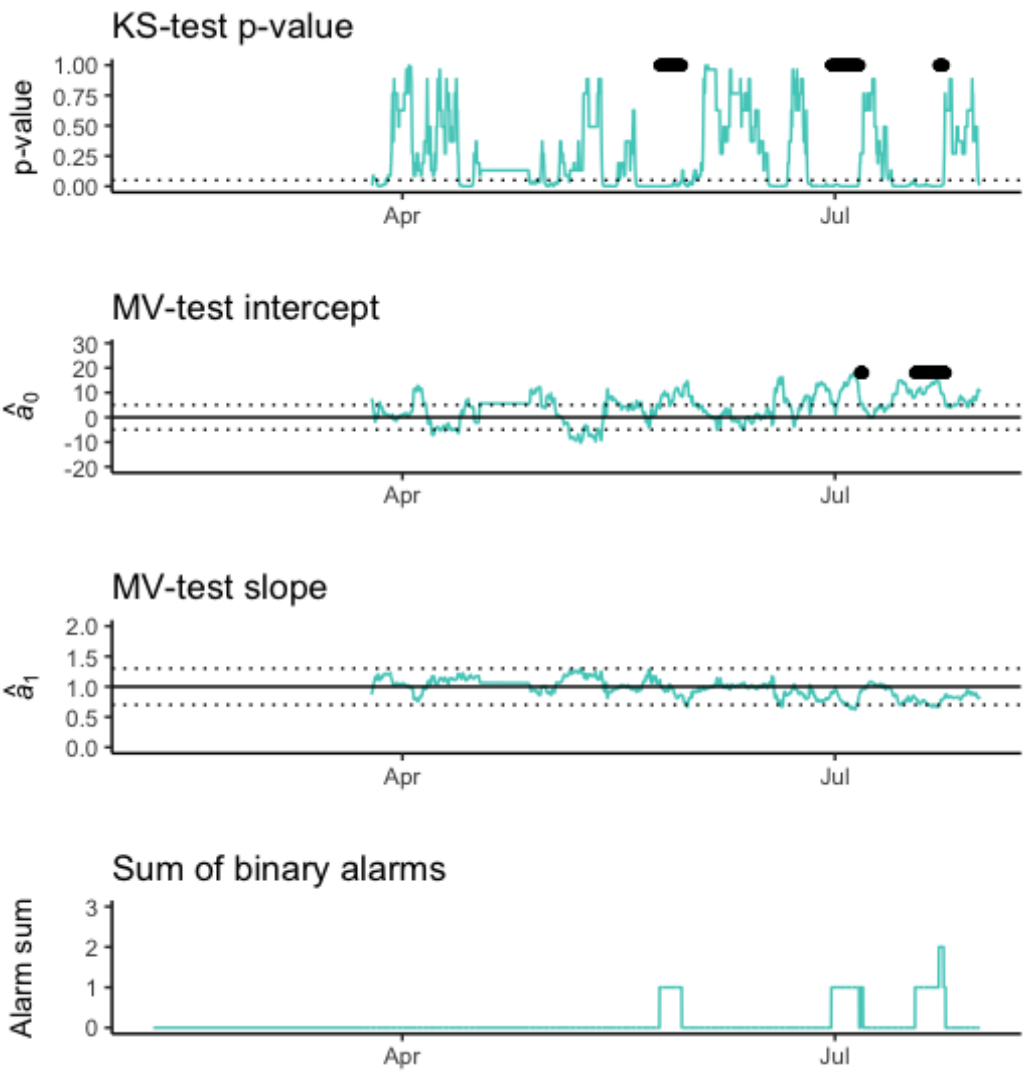

4. Summary results for the application of the framework on the sensor data, using closest other regulatory station as proxy.

| Site ID | Sensor ID | KS $p_{KS}$ alarm % | MV $\hat{a}_0$ % | MV $\hat{a}_1$ % | Framework-corrected % |
|---|---|---|---|---|---|
| RIVR | 100 | 13 | 0 | 0 | 0 |
| MLVB | 101 | 13 | 0 | 0 | 0 |
| SNBO | 102 | 17 | 39 | 19 | 22 |
| FONT | 103 | 9 | 17 | 0 | 0 |
| RDLD | 104 | 17 | 38 | 13 | 18 |
| PICO | 161 | 11 | 3 | 0 | 0.3 |
| CMPT | 166 | 5 | 0 | 0 | 0 |
| LAXH | 176 | 35 | 21 | 0 | 17 |
| HDSN | 177 | 5 | 0 | 0 | 0 |
| CELA | 182 | 11 | 2 | 0 | 0.3 |

The table shows the percentage of the total test time recorded as an alarm by each of the individual tests, and the percentage of the total time that the data were corrected using the framework described in the main text

5. Map showing the movement of "buddy" sensors from regulatory sites to test sites

Map showing the movement of the "buddy" sensors (main text, section 3.4). Names refer to the initial co-located regulatory station. The blue locations where the arrows point to are the locations to which the "buddy" sensors were moved, to check the sensor mounted there. The pink locations are the locations of the regulatory sites where the "buddy" sensors were calibrated and from which they were moved.

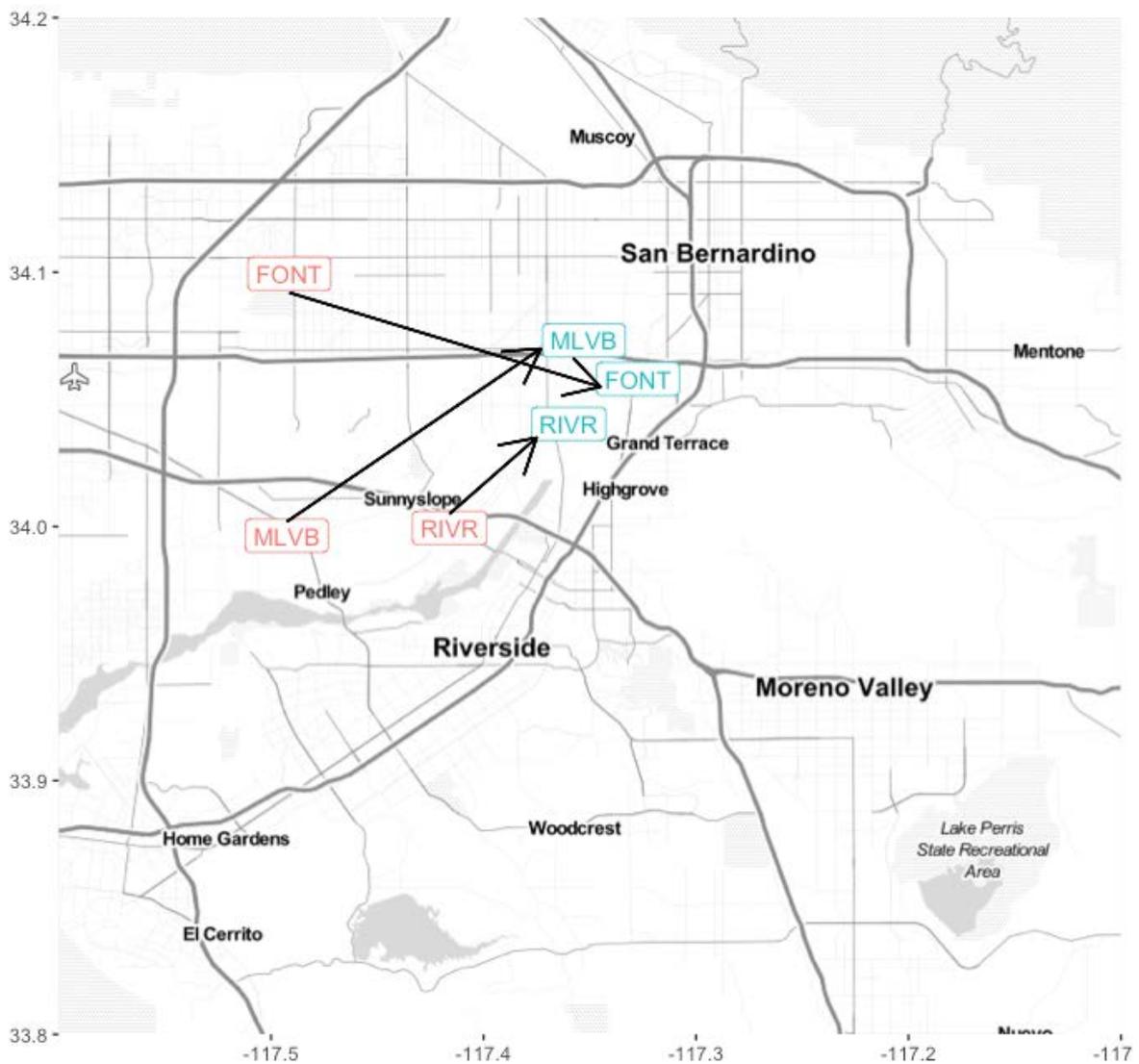